\documentclass[prd,twocolumn,showpacs,floatfix,amsmath,amssymb,floatfix,nofootinbib,showkeys]{revtex4}
\usepackage{graphicx,color,dcolumn,booktabs,bm}
\usepackage{longtable,lscape,comment,braket}
\usepackage{txfonts}
\usepackage{overpic}
\usepackage{amssymb}
\usepackage{indentfirst}
\usepackage{feynmf}   %{feynmp}
\usepackage{slashed}  %for Feynman symbols
\usepackage{cases}
\usepackage{epstopdf}
\usepackage{psfrag}
\usepackage{subfigure}
\pdfoutput=1
\usepackage{color}
\usepackage{multirow}
\graphicspath{{Figures/}} %
\usepackage[colorlinks, citecolor=blue,anchorcolor=red,menucolor=red, linkcolor=red,filecolor=red,runcolor=red,urlcolor=blue,frenchlinks=red]{hyperref}

\def\be{\begin{equation}}
\def\ee{\end{equation}}

\begin{document}
%\begin{CJK}{GBK}{}

\title{Using $X(3823)\to J/\psi\pi^+\pi^-$ to identify coupled-channel effects}
\author{Bo Wang$^{1,2}$}\email{wangb13@lzu.edu.cn}
\author{Hao Xu$^{1,2}$}\email{xuh2013@lzu.cn}
\author{Xiang Liu$^{1,2}$}\email{xiangliu@lzu.edu.cn}
\author{Dian-Yong Chen$^{1,2}$}\email{chendy@impcas.ac.cn}
\author{Susana Coito$^{2}$}\email{susana@impcas.ac.cn}
\author{Estia Eichten$^{5}$}\email{eichten@fnal.gov}
\affiliation{
$^1$Research Center for Hadron and CSR Physics, Lanzhou University and Institute of Modern Physics of CAS, Lanzhou 730000, China\\
$^2$School of Physical Science and Technology, Lanzhou University, Lanzhou 730000, China\\
$^3$Research Center for Hadron and CSR Physics, Lanzhou University and Institute of Modern Physics of CAS, Lanzhou 730000, China\\
$^4$Institute of Modern Physics, Chinese Academy of Sciences, Lanzhou 730000, China\\
$^5$Theoretical Physics Department, Fermilab, IL 60510, USA}

\begin{abstract}
Very recently, the Belle and BESIII experiments observed a new charmonium-like state $X(3823)$, which is a good candidate for the $D$-wave charmonium $\psi(1^3D_2)$.   Because the $X(3823)$ is just near the $D\bar{D}^*$ threshold,  the decay $X(3823)\to J/\psi\pi^+\pi^-$ can be a golden channel to test the significance of coupled-channel effects.   In this work, this decay is considered including both the hidden-charm dipion and the
usual quantum chromodynamics multipole expansion (QCDME) contributions. The partial decay width, the dipion invariant mass spectrum distribution $\mathrm{d}\Gamma[X(3823)\to J/\psi\pi^+\pi^-]/\mathrm{d}m_{\pi^+\pi^-}$, and the corresponding $\mathrm{d}\Gamma[X(3823)\to J/\psi\pi^+\pi^-]/\mathrm{d}\cos\theta$ distribution are computed.   Many parameters are determined from existing experimental data, so the results depend mainly only on one unknown phase
between the QCDME and hidden-charm dipion amplitudes.
\end{abstract}

\maketitle

\section{Introduction}\label{sec1}

Charmonium spectroscopy plays an important role in understanding strong interactions. New states appear continuously in  experiments and are still puzzling, as they seem not to fit the model predictions. See Refs.~\cite{Liu:2013waa,Swanson:2006st,Brambilla:2010cs,Zhu:2007wz,Olsen:2014qna,Yuan:2015kya} for reviews. However, some quark model low-lying states, such as the $\eta_{c2}(1^1D_2)$, $\psi(1^3D_2)$, or even the state $\psi(1^3D_3)$, are still missing \cite{pdg2014}. Observation of the properties of these missing charmonium states can distinguish different phenomenological models, namely, quenched models, which consider only the naive $q\bar{q}$ spectrum (e.g., \cite{Godfrey:1985xj}), and unquenched models, where the coupled-channel effect is considered to be relevant \cite{Eichten:1978tg}.

Experiments show evidence of a state very likely to be interpreted as the $\psi(1^3D_2)$, the $X(3823)$. In 1994, the E705 experiment indicated a state in channel $J/\psi\pi^+\pi^-$ with 2.8$\sigma$ and mass and width $m=3836\pm13$ MeV and $\Gamma=24\pm5$ MeV, respectively \cite{Antoniazzi:1993jz}.
Interpretation as the $1^3D_2$ was favored, since the $1^1D_2$ decays into this channel is suppressed by G-parity, and the $1^3D_3$ decays to the Okubo--Zweig--Iizuka (OZI)-allowed opened channel $D\bar{D}$, making it less likely to be seen in an OZI-suppressed channel. However, the observation of the $X(3823)$ was not confirmed by other experiments in the following 19 years. Finally, in 2013, Belle reported evidence of a new charmonium-like state in the radiative decay to $\chi_{c1}\gamma$ with mass $3823.1\pm 1.8({\rm stat})\pm0.7({\rm syst})$ MeV and significance 3.8 $\sigma$ \cite{Bhardwaj:2013rmw}. Very recently, BESIII confirmed the signal in the $\chi_{c1}\gamma$ invariant mass spectrum with a significance of 6.2$\sigma$ in the process $e^+e^-\to \pi^+\pi^-\gamma\chi_{c1}$, with a measured mass of $3821.7\pm1.3({\rm stat})\pm0.7({\rm syst})$ MeV and a width of less than 16 MeV \cite{Ablikim:2015baa}. Therefore, the $X(3823)$ is now firmly established.

The $X(3823)$ is consistent with the theoretical prediction of the long-missing charmonium $\psi(1^3D_2)$ \cite{Eichten:2002qv}. Although the mass of $\psi(1^3D_2)$ exceeds the $D\bar{D}$ threshold, the $\psi(1^3D_2)\to D\bar{D}$ channel is forbidden by parity conservation. Thus, no OZI-allowed open-charm decay modes exist, so the $\psi(1^3D_2)$ is expected to be a very narrow state. As a triplet state, the $\psi(1^3D_2)$ should typically decay to the $1^3S_1\pi\pi$, alias $J/\psi\pi\pi$, and decay radiatively to $1^3P_1\gamma$ and $1^3P_2\gamma$, alias $\chi_{c1}\gamma$ and $\chi_{c2}\gamma$, respectively. All of these channels have been analyzed in the above experiments, and enhancements have been detected. In addition, the upper limit for the ratio $B(X(3823)\to \chi_{c2}\gamma)/B(X(3823)\to \chi_{c1}\gamma)$ was given as $<0.41$ by Belle and $<0.42$ by BESIII; these values are consistent with prior theoretical calculations in Refs.~\cite{Eichten:2002qv,Ebert:2002pp,Ko:1997rn,Qiao:1996ve}, which give an upper ratio of around $0.24$. In the same works, the partial decay width for $\psi(1^3D_2)\to J/\psi\pi^+\pi^-$ is estimated to be around 45 keV, a value well below the upper limit for the signal observed at E705. All this evidence strongly supports the identification of the $X(3823)$ with the $\psi(1^3D_2)$.
\begin{figure}[hptb]
\setlength{\abovecaptionskip}{-0.1cm}
\setlength{\belowcaptionskip}{-0.1cm}
\begin{center}
\scalebox{1.0}{\includegraphics[width=\columnwidth]{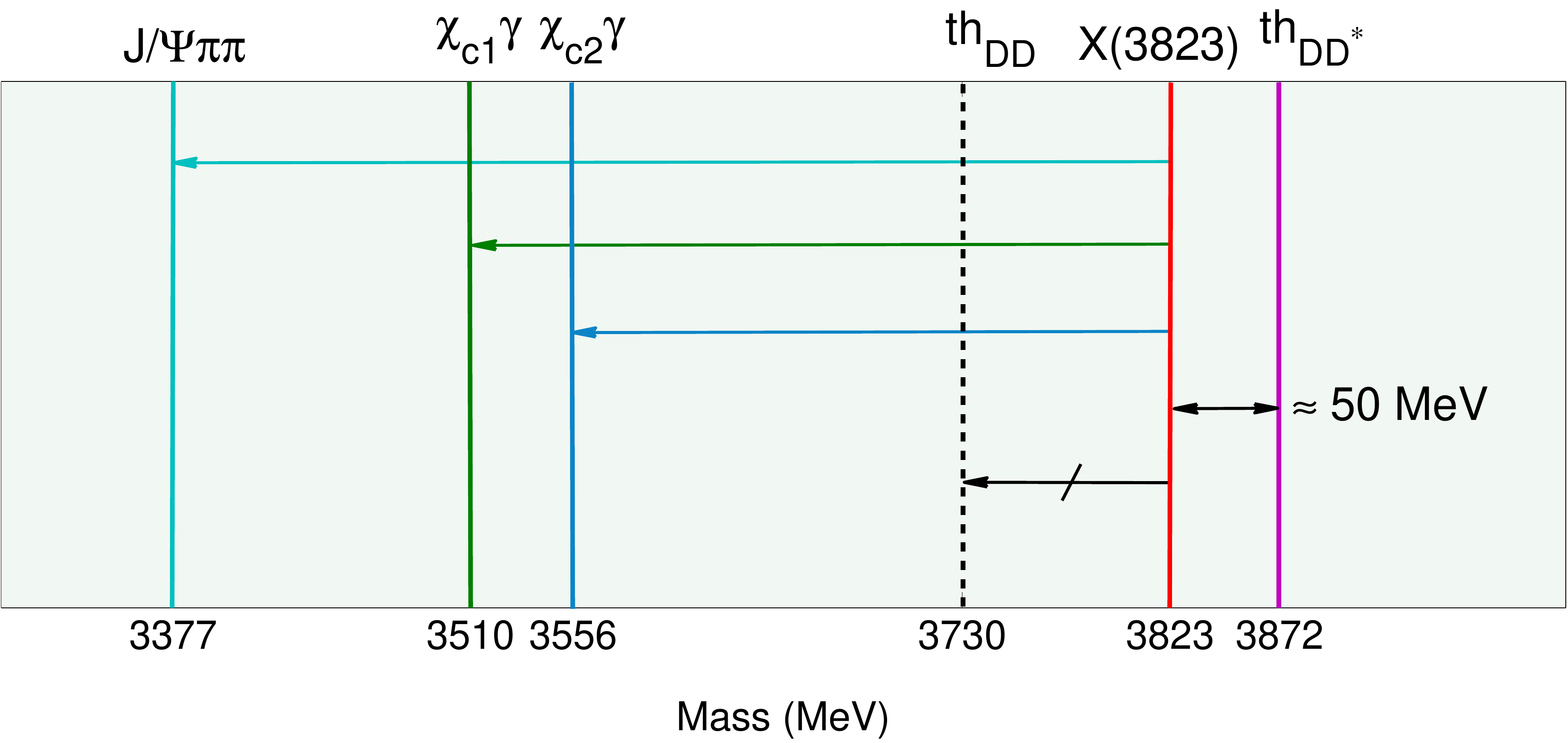}}
\end{center}
\caption{(color online). Comparison of the mass of $X(3823)$ with the $D\bar{D}$ and $D\bar{D}^*$ thresholds, and some allowed and forbidden decay channels.}\label{decaymodes}
\end{figure}

In theory, all OZI-allowed decay channels, which are not forbidden by conservation of quantum numbers, have nonzero coupling to the bare state $q\bar{q}$ even if they are closed. We observe that the $D\bar{D}^{*}$ threshold lies only about $50$ MeV above the mass of the $X(3823)$, as shown in Fig.~\ref{decaymodes}. We suppose that the influence of $D\bar{D}^{*}$ in the $X(3823)$ state may be visible in its strong decay to $J/\psi\pi\pi$. This idea is supported by several studies of hadronic transitions between heavy quarkonia,  suggesting the relevance of coupled-channel effects \cite{Kuang:2006me,Liu:2006dq,Simonov:2007bm,Liu:2009dr,Zhang:2009kr,Chen:2011jp,Chen:2014ccr}.

In the present work, the $X(3823)$ is assumed to be the $\psi(1^3D_2)$; inspired by the observation of the $X(3823)$ so near the  $D\bar{D}^*$  threshold,  we investigate the influence of the $D\bar{D}^*$ channel in the hidden-charm dipion decay $X(3823)\to J/\psi\pi^+\pi^-$. For this purpose, we study the dipion invariant mass and the scattering angle distributions of this process with and without the influence of $D\bar{D}^*$. Moreover, we try to compare the results with the sparse data from the E705 experiment. We employ two different methods. First, we apply the quantum chromodynamics multipole expansion (QCDME) \cite{Kuang:1988bz} to study the decay $X(3823)\to J/\psi\pi^+\pi^-$ without the coupled-channel effects and obtain the distribution of the dipion invariant mass.
Second, we calculate the same process including the hadronic loop mechanism, which is an effective description of the coupled-channel effect \cite{Liu:2006dq}.
By adding $D\bar{D}^*$, we illustrate the change in the distributions of the dipion invariant mass and the polar angle with the interference of the coupled channel. Because these results might be accessible in current experiments, we consider that $X(3823)\to J/\psi\pi^+\pi^-$ is a golden channel to test the coupled-channel effect, and we also expect to motivate further analysis. We would like to note the relevance of this study to future theoretical development. The $\psi(1^3D_2)$ is a charmonium state, i.e., a heavy quark state, below all possible OZI-allowed thresholds. If we can show that coupling to a not-so-nearby closed channel is relevant to a faithful description of the state, this means that we surely cannot neglect the influence of the OZI-allowed decay channels in any serious description of resonances or light-quark systems \cite{Rupp:2015}.

This paper is organized as follows. In Sec.~\ref{sec2}, we describe the study of $X(3823)\to J/\psi\pi^+\pi^-$ via the QCDME method. In Sec.~\ref{sec3}, we present the calculation details of the same process using the hadronic loop mechanism, and in Sec.~\ref{sec4}, we show numerical results for the dipion invariant mass and scattering angle including the coupled-channel effects. The paper ends with a summary in Sec.~\ref{sec5}.

\section{Study of $X(3823)\to J/\psi\pi^+\pi^-$ without coupled-channel effects}\label{sec2}

In this section, we employ the QCDME method to study the hadronic transition $X(3823)\to J/\psi\pi^+\pi^-$ without considering any coupled-channel effect. The method has typically been applied to OZI-suppressed decays of heavy quarkonia with emission of light quarks and has proved to be reliable for general predictions of $\pi\pi$ invariant mass distributions and decay widths. The general idea is that a heavy quarkonium de-excites to a lower energy level, i.e., a lower radial level or orbital angular momentum level, by radiating gluons, in a manner very similar to that of the electromagnetic transitions within an atom. However, in contrast to the electromagnetic case, two complex vertices are involved; in the first vertex, we have multipole gluon emissions, whereas in the second vertex, hadronization occurs. Details of the QCDME method can be found in Refs. \cite{Kuang:2006me,Yan:1980uh,Kuang:1981se}.

Our process is a transition between two triplet states, $1^3D_2\to1^3S_1\pi\pi$, which is dominated by double color-electric dipole emissions (E1-E1). By identifying $\Phi_i,\ \Phi_f$, and $h$ with the initial and final heavy quarkonium $Q\bar{Q}$ states and the emitted light hadrons, respectively, the transition amplitude can be expressed as \cite{Kuang:2006me,Yan:1980uh}
 \begin{eqnarray}
 \mathcal{M}_{{E1-E1}}=i\frac{g^2_E}{6}\left\langle\Phi_f h\left|\vec{x}\cdot\vec{E}G(E_i)\vec{x}\cdot\vec{E}\right|\Phi_i\right\rangle, \label{appd4}
 \end{eqnarray}
where $g_E$ is the effective coupling constant for the chromo-electric multipole gluon emissions, $\vec{x}$ is the separation between $Q$ and $\bar{Q}$, and $\vec{E}$ is the color-electric field. The Green function $G(E_i)$, where $E_i$ is the initial state energy, is given by
  \begin{eqnarray}\label{gf}
 G(E_i)=1/(E_i-H_8-iD_0)
  \end{eqnarray}
with the gauge covariant time derivative
  \begin{eqnarray}\label{cd}
 D_0=\partial_0-g\bar{A}_0.
   \end{eqnarray}
Equations \eqref{gf} and \eqref{cd} represent propagation of the intermediate states between the two color-electric dipole vertices. In these vertices, we have color-singlet states composed of the $Q\bar{Q}$ color octet, gluons, and light quarks. The gluon field is represented by $\bar{A}_0$, and $H_8$ denotes the octet component of the Hamiltonian.

Equation \eqref{appd4} must be simplified to a calculable form. Because we do not understand the confinement mechanism, we need to introduce two further models to account for the unknown variables at both vertices. For the first vertex, we assume the quark-confining string model. Here, the intermediate states, i.e., those after emission of the first gluon $g$ and before emission of the second gluon, are considered to be hybrid states. The ground state is simply a string between $Q$ and $\bar{Q}$, and the first vibrational mode is the hybrid $Q\bar{Q}g$, the only one we keep. For the hadronization vertex, the soft-pion theorem is assumed.

After manipulating Eq.~\eqref{appd4}, we find that the transition rate $\Gamma(\Phi_i\to\Phi_f\pi\pi)$ between spin triplets with $l_i=2$ and $l_f=0$, where $l_{i,f}$ is the orbital momentum of the initial and final states, respectively, gives \cite{Kuang:1981se}
\begin{eqnarray}
\Gamma[^3D\to\ ^3S \pi^+\pi^-]=\frac{4}{15}\mathcal{H}\,|c_2|^2\,|f_{if}^1|^2,\label{eqn1}
\end{eqnarray}
where $\mathcal{H}$ denotes the phase-space integral:
\begin{eqnarray}\label{eqh}
\mathcal{H}&=&\frac{\pi^3 m_{J/\psi}}{20m_{X}}\int dm_{\pi\pi}^2\mathcal{K}\left(1-\frac{4m_\pi^2}{m_{\pi\pi}^2}\right)^{1/2}\Bigg[\Big(m_{\pi\pi}^2-4m_\pi^2\Big)^2\nonumber\\
&&\times\left(1+\frac{2}{3}\frac{\mathcal{K}^2}{m_{\pi\pi}^2}\right)+\frac{8\mathcal{K}^4}{15m_{\pi\pi}^4}\Big(m_{\pi\pi}^4+2m_\pi^2m_{\pi\pi}^2+6m_\pi^4\Big)\Bigg],\label{appd1}
\end{eqnarray}
where
\begin{eqnarray}
\mathcal{K}=\frac{1}{2m_X}\Big[(m_X+m_{J/\psi})^2-m_{\pi\pi}^2\Big]^{1/2}\Big[(m_X-m_{J/\psi})^2-m_{\pi\pi}^2\Big]^{1/2}\label{appd2}.
\end{eqnarray}
The dynamical part $f_{if}^1$ is expressed as
\begin{eqnarray}\label{eqf}
f_{if}^1&=&\sum_n\frac{1}{m_i-m_{n1}}\left[\int drr^3\mathcal{R}_f(r)\mathcal{R}_{n1}(r)\right]\nonumber\\
&&\times\left[\int dr^\prime r^{\prime3}\mathcal{R}_{n1}(r^\prime)\mathcal{R}_i(r^\prime)\right]\label{appd3},
\end{eqnarray}
where $\mathcal{R}_i(r)$, $\mathcal{R}_f(r)$, and $\mathcal{R}_{n1}(r)$ are the radial wave functions of the initial, final, and intermediate vibrational states, respectively, and the subscripts $1$ and $n$ correspond to the orbital angular momentum and radial quantum number, respectively. The radial wave functions are obtained numerically by solving the Schr\"{o}dinger equation using the Cornell potential, which is defined as
\begin{eqnarray}\label{funnel}
V(r)=\frac{r}{a^2}-\frac{\kappa}{r},
\end{eqnarray}
where $a=2.34 \text{ GeV}^{-1}$ and $\kappa=0.52$, and the constituent charm quark mass $m_c$ is 1.84 GeV \cite{Kuang:1981se}.

{To obtain the radial wave function $R_{n1}$ of the intermediate vibrational states, we introduce the potential model given in Ref. \cite{Buchmuller:1979gy}:
\begin{eqnarray}
V_\nu(r)=V(r)+\left[V_n(r)-\frac{1}{a^2}r\right]+\frac{A_\nu}{r}.\label{funne2}
\end{eqnarray}
Here, $V(r)$ is listed in Eq.~(\ref{funnel}), and $V_n(r)$ is given by
\begin{eqnarray}
V_n(r)&=&\frac{1}{a^2}r\left[1+\frac{2\pi a^2n}{(r-2d)^2+4d^2}\right]^{1/2}\nonumber\\
&\equiv&\frac{1}{a^2}r[2-\alpha_n^2(r)]^{-1/2},\nonumber\\
d&=&\frac{r^2\alpha_n(r)}{4a^2[2m_c+(1/a^2)r\alpha_n(r)]}.\label{funne3}
\end{eqnarray}
We consider only the lowest string excitation, which corresponds to $n=1$ in Eq.~(\ref{funne3}), and adjust the constant $A_\nu$ to fit the mass of the lowest vibrational state. As treated in Ref. \cite{Kuang:1981se}, $m_\nu=4.03$ GeV is taken as the mass of the ground state of the $c\bar{c}$ vibrational spectrum.
}

%{\color{red}  You never define how you determine $R_{nl}$. Put a few words here. EE}
To determine the unknown parameter $c_2$ in Eq.~\eqref{eqn1}, we use our knowledge of the decays $\psi(3686)\to J/\psi\pi^+\pi^-$ and $\psi(3770)\to J/\psi\pi^+\pi^-$, which are both transitions between triplet states. Here, we consider the mixing
\be
\begin{split}
&\psi(3686)=\psi_{2S}\cos\theta+\psi_{1D}\sin\theta,\\
&\psi(3770)=-\psi_{2S}\sin\theta+\psi_{1D}\cos\theta.
\end{split}
\ee
From the nonrelativistic formulas for the leptonic decay widths and the experimental values for the decays $\psi(3686),\psi(3770)\to e^+e^-$, the mixing angle is found to be around $-10^\circ$. Moreover, the decay amplitude for $\psi(3686),\psi(3770)$ involves a mixture between Eq. \eqref{eqn1} and a similar expression for the decay $\Gamma(^3S\to\ ^3S \pi\pi)$, which involves a new parameter, $c_1$. However, as there are accurate data for the transitions $\psi(3686),\psi(3770)\to J/\psi\pi\pi$, both parameters, $c_1$ and $c_2$, are fully determined. We obtain $|c_2|^2\simeq1.46\times10^{-4}$ for the partial width in  $\psi(^3D_2)\to J/\psi\pi^+\pi^-$.

For the process $X(3823)\to J/\psi\pi^+\pi^-$, Eqs.~\eqref{eqh}--\eqref{funnel} give the results $\mathcal{H}=0.0176$ GeV$^7$ and $f_{if}^1=-11.4$ GeV$^{-3}$. Finally, the partial decay width becomes
\begin{eqnarray} \label{decw}
 \Gamma[X(3823)\to J/\psi\pi^+\pi^-]\simeq89.1 \text{ keV},
\end{eqnarray}
which is about two times larger than the previous result in Ref.~\cite{Eichten:2002qv}.\\

\begin{figure}[hptb]
\setlength{\abovecaptionskip}{-0.03cm}
\scalebox{1.0}{\includegraphics[width=\columnwidth]{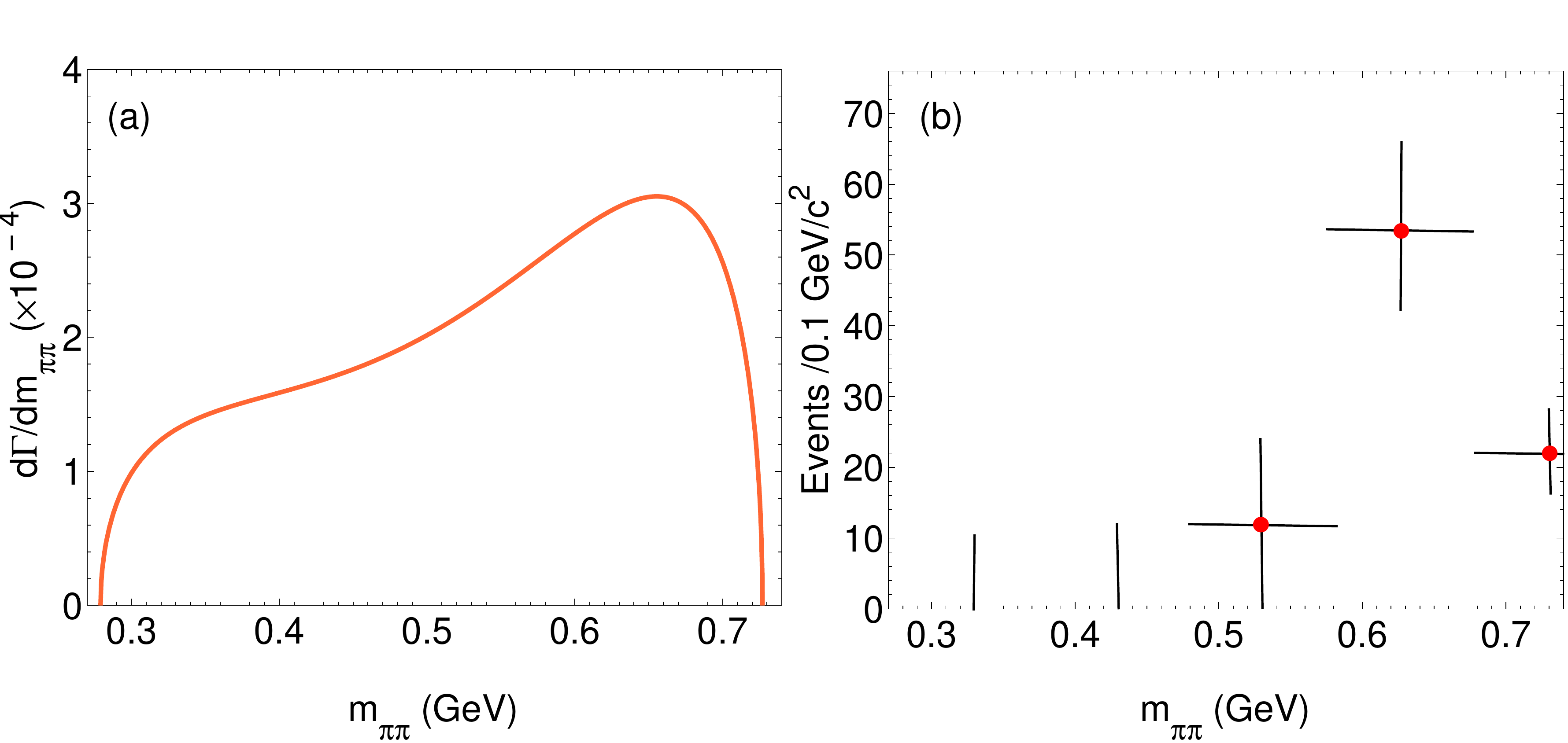}}
\caption{(color online). The $\pi^+\pi^-$ invariant mass distribution $d\Gamma[X(3823)\to J/\psi\pi^+\pi^-]/dm_{\pi^+\pi^-}$ calculated using the QCDME method (left-hand panel) and the experimental data from E705 \cite{Antoniazzi:1993jz} (right-hand panel).}\label{qcdspectrum}
\end{figure}

We first consider the $\Gamma(X(3823)\to J/\psi\pi^+\pi^-)$ distribution over the $\pi^+\pi^-$ invariant mass, commonly called the dipion invariant mass distribution for simplicity. The result is shown in Fig.~\ref{qcdspectrum} (a). The lower $m_{\pi\pi}$ kinematic region is dominated by the $S$-wave contribution in the $\pi^+\pi^-$ propagation, whereas the peak around 0.65 GeV in the $\pi^+\pi^-$ invariant mass spectrum is due to the $D$-wave contribution \cite{Voloshin:2015jua}. We also show the corresponding experimental data from the E705 \cite{Antoniazzi:1993jz} experiment in Fig.~\ref{qcdspectrum} (b). A comparison of our QCDME results with the experimental data indicates a clear  discrepancy in the $m_{\pi^+\pi^-}$ low-momentum region. This result is somewhat expected because more nonperturbative effects should be revealed at lower energies. Because QCDME appears to be insufficient to describe the data, we need to consider a new effect on $X(3823)\to J/\psi\pi^+\pi^-$. As mentioned in Sec.~\ref{sec1}, the proximity of $X(3823)$ to the closed $D\bar{D}^*$ threshold inspires our interest in studying the coupled-channel effects in the decay. This will be the task in the next section.

%In the following, we make a rough estimation about the total width of $X(3823)$, where $X(3823)$ is expected to be a narrow state since the OZI-allowed hadronic decay channel $D\bar{D}$ is forbidden by the parity conservation \cite{Bhardwaj:2013rmw,Ablikim:2015baa}. The possible decay channels of $X(3823)$ include $J/\psi\pi\pi$, $\chi_{c1}\gamma$, $\chi_{c2}\gamma$ and $ggg$, for the decay magnitudes of $X(3823)$ into $\chi_{c1}\gamma$, $\chi_{c2}\gamma$ and $ggg$ we adopt the values in Ref. \cite{Eichten:2002qv}. Therefore, $\Gamma_{\mathrm{Total}}[X(3823)]\approx\Gamma[X(3823)\to J/\psi\pi\pi]+\Gamma[X(3823)\to \chi_{c1}\gamma]+\Gamma[X(3823)\to \chi_{c2}\gamma]+\Gamma[X(3823)\to ggg]\approx$ 478 keV, which is consistent with the experimental measurements for the upper limit of total width \cite{Bhardwaj:2013rmw,Ablikim:2015baa}.

\section{Including the  coupled-channel effects}\label{sec3}
%\subsection{formulism}

\begin{figure}[hptb]
\scalebox{0.97}{\includegraphics[width=\columnwidth]{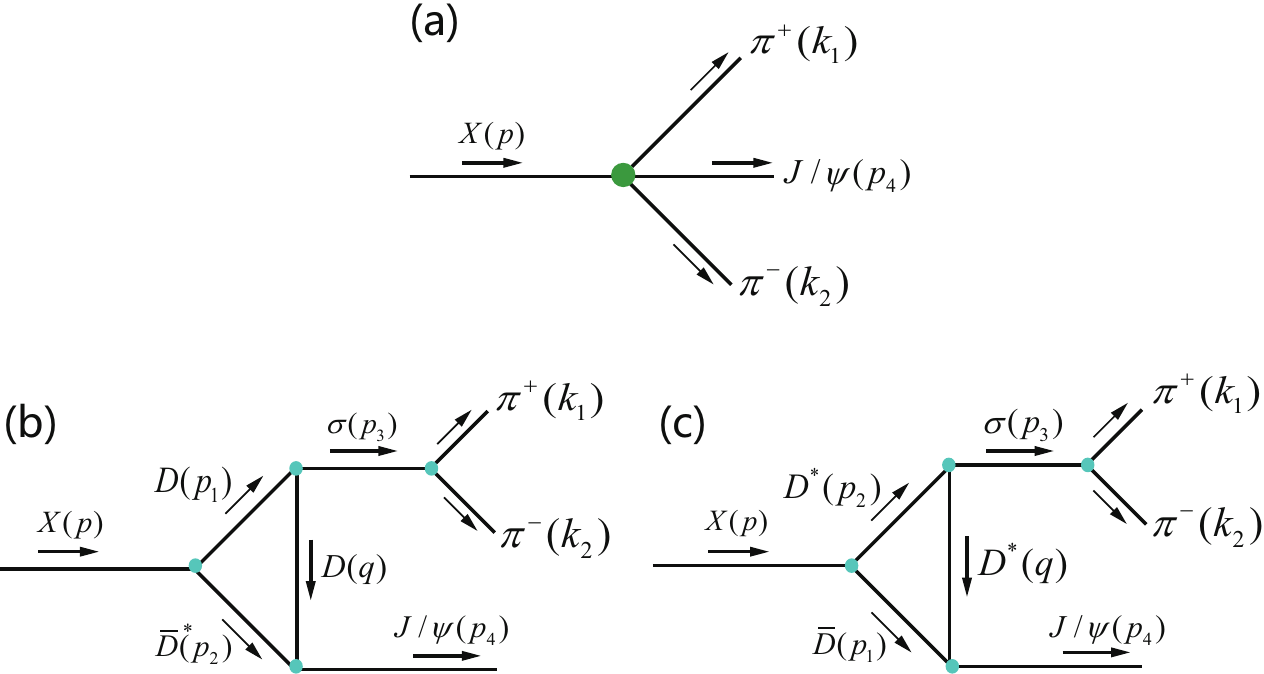}}
\caption{(color online). Diagrams describing $X(3823)\to J/\psi\pi^+\pi^-$. Diagram (a) represents the direct decay process without considering the coupled-channel effects, whereas diagrams (b) and (c) show the coupled-channel contribution.\label{feynman}}
\end{figure}

In this section, we adopt the effective Lagrangian approach, which was previously used for several other quarkonium systems  \cite{Chen:2011zv,Chen:2011qx,Chen:2011xk,Meng:2007cx}, to study the decay $X(3823)\to J/\psi\pi^+\pi^-$ by including hadronic loops, i.e., coupled-channel effects.
Figure~\ref{feynman} shows diagrams illustrating the processes on the hadron level, where the triangle diagrams in (b) and (c) represent the coupled-channel contribution. Here, the decay $X(3823)\to J/\psi\pi^+\pi^-$ occurs in two steps, $X(3823)\to D\bar{D}^*+h.c.$, followed by (b) $D\bar{D}\to\sigma\to\pi^+\pi^-$ or (c) $D^*\bar{D}^*\to\sigma\to\pi^+\pi^-$, and $D\bar{D}^*+h.c.\to J/\psi$.

%{\color{red} remove this paragraph??}
%We need to indicate that indeed there are additional diagrams with the emission of two independent pions from some combination of the vertices of the initial and final states and the $D$-meson loops. It would be expected that these contributions are higher order in the chiral expansion and would subdominant to the emission of a single  $\sigma$ meson.
%The inclusion of these diagrams would also introduce addition free introduced parameters like (e.g. an extra phase) which makes whole study more complicated and uncontrolled.  To establish the importance of the coupled-channel effects, we only adopted the sigma meson as the intermediate contribution to $X(3823)\to J/\psi \pi^+\pi^-$. {In this paper, we assume a direct dominate coupling of the sigma meson to $D$ mesons.} Of course,  it would be useful to also study the two independent pion contributions when more data of these decays are available.
%At present the experimental data of $X(3823)\to J/\psi \pi^+\pi^-$ are scarce. If more experimental data will be available by future experiment, we will include these extra diagrams to do a global fit to the experimental data, which will be an intriguing research topic.

For the direct diagram in Fig.~\ref{feynman} (a), {we adopt the method in Refs.~ \cite{Voloshin:2015jua,Voloshin:1980zf,Voloshin:2006ce}, which are also based on the concepts of QCDME. The conversion of gluons into two pions is described by the matrix element $\langle\pi^+(k_1)\pi^-(k_2)|g^2F_{\mu\nu}^a(x)F_{\lambda\sigma}^a(x)|0\rangle$, where $g$ is the QCD coupling constant, $a$ is the color index, and $F_{\mu\nu}^a(x)$ is the gluon field strength operator. This matrix element can be uniquely determined using triangle anomalies and the trace of the energy-momentum tensor, as well as the soft pion approximation (see Refs. \cite{Voloshin:1980zf,Voloshin:2006ce} for more detailed derivations). Voloshin \cite{Voloshin:2015jua,Voloshin:2006ce} constructed the amplitude of the $\pi\pi$ transition between heavy quarkonium and explicitly separated the $S$- and $D$-wave contributions in the matrix element. For the direct diagram shown in Fig.~\ref{feynman} (a), the decay amplitude is \cite{Voloshin:2015jua,Voloshin:2006ce}}
\begin{eqnarray}
\mathcal{M}^{(a)}=ig_{XJ/\psi\pi\pi}\epsilon_{\kappa\lambda\nu\sigma}ip_4^\kappa\epsilon_{J/\psi}^{\ast\lambda} \epsilon^\nu_{X\mu}\left[\frac{2}{3}\Big(1+\frac{2m_\pi^2}{p_3^2}\Big)p_3^\sigma p_3^\mu-\ell^{\sigma\mu}\right].\label{eqn13}
\end{eqnarray}
with
\begin{eqnarray}
 \ell^{\sigma\mu}=\tilde{q}^\sigma \tilde{q}^\mu+\frac{1}{3}\Big(1-4m_\pi^2/p_3^2\Big)(p_3^2g^{\sigma\mu}-p_3^\sigma p_3^\mu),
 \end{eqnarray}
where $g_{XJ/\psi\pi\pi}$ is the partial coupling constant corresponding to the direct process $X(3823)\to J/\psi\pi^+\pi^-$, $\epsilon_{\kappa\lambda\nu\sigma}$ is the antisymmetric symbol, and { $\epsilon_{J/\psi}^{\ast\lambda}$ and $\epsilon^\nu_{X\mu}$ correspond to the polarization vector of $J/\psi$ and polarization tensor of $X(3823)$}, respectively. The first and second terms are defined as $p_3=k_1+k_2$ and $\tilde{q}=k_1-k_2$, where $k_1,\ k_2$ corresponds to the $\pi^+\pi^-$ 4-momentum; these terms represent the $S$- and $D$-wave contributions to dipion propagation, respectively.

For the diagrams with hadron loops in Fig.~\ref{feynman} (b) and (c), we use the heavy quark effective model. For a heavy-light meson system, in the heavy quark limit $m_Q\to\infty$, there is heavy quark spin symmetry and heavy quark flavor symmetry. However, for a heavy quarkonium such as charmonium, heavy quark flavor symmetry will not hold while heavy quark spin symmetry still exists \cite{Casalbuoni:1996pg}. Thus, for charmonium systems, states with a given orbital angular momentum $L$ but different spins form a multiplet. For example, the $L=0$ ($S$-wave) charmonium spin doublet $\mathcal{J}$ can be written as \cite{Colangelo:2003sa,Casalbuoni:1992fd}
\begin{eqnarray}
\mathcal{J}=\frac{1+\slashed{v}}{2}\Big[J/\psi^\mu\gamma_\mu-\eta_c\gamma_5\Big]\frac{1-\slashed{v}}{2},\label{l0}
\end{eqnarray}
where $v^\mu$ is the 4-velocity of the multiplet, and $J/\psi^\mu$ and $\eta_c$ are the spin-1 and spin-0 components, respectively.
The general form of the orbital angular momentum $L\neq0$ multiplet has been established in Refs. \cite{Casalbuoni:1992fd,Casalbuoni:1996pg}. For $L=2$, the charmonium multiplet is given by
\begin{eqnarray}
\mathcal{J}^{\mu\lambda}&=&\frac{1+\slashed{v}}{2}\Bigg[X_3^{\mu\lambda\alpha}\gamma_\alpha+\frac{1}{\sqrt{6}}\Big(\epsilon^{\mu\alpha\beta\rho}v_\alpha\gamma_\beta X_{\rho}^\lambda+\epsilon^{\lambda\alpha\beta\rho}v_\alpha\gamma_\beta X_{\rho}^\mu\Big)\nonumber\\
&&+\frac{\sqrt{15}}{10}\Big[(\gamma^\mu-v^\mu)\psi^\lambda+(\gamma^\lambda-v^\lambda)\psi^\mu\Big]\nonumber\\
&&-\frac{1}{\sqrt{15}}\Big(g^{\mu\lambda}-v^\mu v^\lambda\Big)\gamma_\alpha\psi^\alpha+\eta_2^{\mu\lambda}\gamma_5\Bigg]\frac{1-\slashed{v}}{2}, \label{eqn3}
\end{eqnarray}
where all the tensor fields are traceless, symmetric, and transverse. The fields $X_3$, $X$, $\psi$, and $\eta_2$ denote the charmonia with $J^{PC}=3^{--}$, $2^{--}$, $1^{--}$, and $2^{-+}$, respectively, where $X$ and $\psi$ correspond to $X(3823)$ and $\psi(3770)$, respectively.

The effective Lagrangian describing the open charm mesons interacting with the $S$- and $D$-wave charmonium multiplet is given by
\begin{eqnarray} \label{XHH}
\mathcal{L} &=&g^\prime \mathrm{Tr}\Big[\mathcal{J}\bar{H}_{2}\overleftrightarrow{\partial}_\mu\gamma^\mu\bar{H}_{1}\Big]+\mathrm{H.c.},\nonumber\\
\mathcal{L} &=& g \mathrm{Tr}\Big[\mathcal{J}^{\mu\lambda}\bar{H}_{2}\overleftrightarrow{\partial}_\mu\gamma_\lambda\bar{H}_{1}\Big]+\mathrm{H.c.}, \label{eqn4}
\end{eqnarray}

where $\overleftrightarrow{\partial}=\overrightarrow{\partial}-\overleftarrow{\partial}$. The mesons with a single heavy quark are represented by $H_{1,2}$, which is defined as
\begin{eqnarray}
H_{1}&=&\frac{1+\slashed{v}}{2}\left[D^{\ast\mu}\gamma_\mu-D\gamma_5\right] \label{h1},\\
H_{2}&=&\left[\bar{D}^{\ast\mu}\gamma_\mu-\bar{D}\gamma_5\right]\frac{1-\slashed{v}}{2}\label{h2},
\end{eqnarray}
and form a doublet with $l=0$, $J^P=(0^-,1^-)$. The field $D^{(\ast)}$ includes a normalization factor $\sqrt{m_{D^{(\ast)}}}$, and $\bar{H}_{1,2}=\gamma^0 H_{1,2}^\dagger\gamma^0$.

Using Eqs.~\eqref{eqn3}--\eqref{h2}, we obtain an explicit expression for the Lagrangian density of $X(3823)$ and $\psi(3770)$ coupling to the charmed meson pair:
\begin{eqnarray}
\mathcal{L}_{XDD^\ast}&=& ig_{XD^\ast D}X^{\mu\nu}\left[\bar{D}\overleftrightarrow{\partial}_\nu D^{\ast}_\mu-\bar{D}^\ast_\mu\overleftrightarrow{\partial}_\nu D\right], \label{eqn5}\\
\mathcal{L}_{\psi DD}&=& g_{\psi DD}\psi^\mu(\bar{D}\partial_\mu D-D\partial_\mu \bar{D}),\label{eqn6}
\end{eqnarray}
where
\begin{eqnarray}
g_{XD^\ast D}&=& 2g\sqrt{\frac{3}{2}}\sqrt{m_X m_{D^\ast} m_D}, \label{XDDstar}\\
g_{\psi DD}&=& -2g\frac{\sqrt{15}}{3}\sqrt{m_\psi}m_D\label{psiDD} .
\end{eqnarray}

For the vertices in Fig.~\ref{feynman} (b) and (c) involving $J/\psi$, we use Eqs.~\eqref{l0} and \eqref{XHH}-\eqref{h2}, which yield
\begin{eqnarray}
\mathcal{L}_{J/\psi DD^\ast}=ig_{J/\psi DD^\ast}\epsilon_{\alpha\rho\beta\lambda}[\bar{D}\overleftrightarrow{\partial}^\alpha D^{\ast\rho}-\bar{D}^{\ast\rho}\overleftrightarrow{\partial}^\alpha D]{\partial}^\beta J/\psi^\lambda.\label{eqn7}
\end{eqnarray}
The effective Lagrangians for the other vertices are given by Refs.
\cite{Chen:2011zv,Chen:2011qx,Chen:2011xk,Meng:2007cx}.

\begin{eqnarray}
\mathcal{L}_{\sigma DD}&=&-g_{\sigma DD}D\bar{D}\sigma,\\\label{eqn8}
\mathcal{L}_{\sigma D^\ast D^\ast}&=&g_{\sigma D^\ast D^\ast}D^\ast\cdot\bar{D}^\ast\sigma,\\\label{eqn9}
\mathcal{L}_{\sigma\pi\pi}&=&g_{\sigma\pi\pi}\sigma\pi\pi.\label{eqn10}
\end{eqnarray}

Given the above Lagrangian densities, the decay amplitudes corresponding to the triangle diagrams in Fig.~\ref{feynman}, with the $p_i$ 4-momentum defined in the figure, are
\begin{eqnarray}
\mathcal{M}^{(b)}+\mathcal{M}^{(c)}=\left[\mathcal{M}_{D\bar{D}^\ast}^D+\mathcal{M}_{\bar{D}D^\ast}^{D^\ast}\right]\times\frac{\sqrt{2}g_{\sigma\pi\pi}}{p_3^2-m_\sigma^2+im_\sigma\Gamma_\sigma},\label{eqn14}
\end{eqnarray}
where
\begin{eqnarray}
\mathcal{M}_{D\bar{D}^\ast}^D&=&(i)^3\int\frac{d^4q}{(2\pi)^4}\Big[ig_{XDD^\ast}\epsilon_{X\mu}^\nu(ip_{2\nu}-ip_{1\nu})\Big] \Big[-g_{\sigma DD}\Big]\nonumber\\
&&\times\Big[ig_{J/\psi DD^\ast}\epsilon_{\alpha\rho\beta\lambda}(iq^\alpha-ip_2^\alpha)ip_4^\beta\epsilon_{J/\psi}^{\ast\lambda}\Big]\frac{1}{p_1^2-m_D^2}\nonumber\\
&&\times\frac{1}{q^2-m_D^2}\frac{-g^{\mu\rho}+p_2^\mu p_2^\rho/m_{D^\ast}^2}{p_2^2-m_{D^\ast}^2}\mathcal{F}^2(q^2), \label{eqn11}
\end{eqnarray}
\begin{eqnarray}
\mathcal{M}_{\bar{D}D^\ast}^{D^\ast}&=&(i)^3 \int\frac{d^4q}{(2\pi)^4}\Big[ig_{XDD^\ast}\epsilon_{X\mu}^\nu(ip_{2\nu}-ip_{1\nu})\Big] \Big[g_{\sigma D^\ast D^\ast}\Big]\nonumber\\
&&\times\Big[ig_{J/\psi DD^\ast}\epsilon_{\alpha\rho\beta\lambda}(-iq^\alpha+ip_1^\alpha)ip_4^\beta\epsilon_{J/\psi}^{\ast\lambda}\Big]\frac{1}{p_1^2-m_D^2}\nonumber\\
&&\times\frac{-g^{\mu\tau}+p_2^\mu p_2^\tau/m_{D^\ast}^2}{p_2^2-m_{D^\ast}^2}\frac{-g^{\rho\tau}+q^\rho q^\tau/m_{D^\ast}^2}{q^2-m_{D^\ast}^2}\mathcal{F}^2(q^2).\nonumber\\   \label{eqn12}
\end{eqnarray}

In Eqs.~\eqref{eqn11} and \eqref{eqn12}, we introduce the monopole form factor
\be
\label{ff}
\mathcal{F}(q^2)=(m_{E}^2-\Lambda^2)/(q^2-\Lambda^2),\ \  \Lambda=m_E+\alpha\Lambda_{QCD}
\ee
to account for the unknown structure and the significant off-shell effect of the exchanged $D^{(\ast)}$ mesons \cite{Liu:2006df,Liu:2006dq}, where $m_E$ and $q$ denote the mass and 4-momentum of the exchanged $D^{(\ast)}$ mesons, respectively, and $\Lambda_{QCD}=220$ MeV.
In Eq.~\eqref{eqn14}, we adopt the momentum-dependent form of $\Gamma_\sigma$ for the propagator of the $\sigma$ meson \cite{Chen:2013coa} because the total decay width and mass are of the same order, i.e.,
\begin{eqnarray}
\Gamma_\sigma(m_{\pi^+\pi^-})=\Gamma_\sigma\frac{m_{\sigma}}{m_{\pi^+\pi^-}}\frac{|\vec{p}(m_{\pi^+\pi^-})|}{|\vec{p}(m_\sigma)|}, \label{eqn15}
\end{eqnarray}
where $|\vec{p}(m_{\pi^+\pi^-})|=\sqrt{m_{\pi^+\pi^-}^2/4-m_\pi^2}$ is the pion momentum, and $|\vec{p}(m_\sigma)|$ is the pion momentum with an on-shell $\sigma$ meson.
After the loop integrals in Eqs.~\eqref{eqn11} and \eqref{eqn12} are performed,
the decay amplitudes in Eq.~\eqref{eqn14} can be further parameterized as
\begin{eqnarray}
\mathcal{M}^{(b)}+\mathcal{M}^{(c)}%&&%=\mathcal{M}[X(3823)\to D^{(\ast)}\bar{D}^{(\ast)}\to J/\psi \pi^+\pi^-]_\sigma\nonumber\\
=4\mathcal{A}p_{3\theta}p_{3\nu}p_{4\eta}\epsilon^{\lambda\mu\theta\eta}\times\frac{\sqrt{2}g_{\sigma\pi\pi}(\epsilon_{X\mu}^\nu\epsilon_{J/\psi\lambda}^\ast)}{p_3^2-m_\sigma^2+im_\sigma\Gamma_\sigma},
\label{eqn16}
\end{eqnarray}
where the amplitudes can be contracted to one independent Lorentz structure, and all the factors are included in $\mathcal{A}$. The factor $4$ comes from the contributions of four possible intermediate channels: $D^0\bar{D}^{\ast0}$, $\bar{D}^0D^{\ast0}$, $D^+D^{\ast-}$, and $D^-D^{\ast+}$.  {Further, $\sqrt{2}$ is the isospin factor of $\pi^+$ and $\pi^-$.}

Finally, the total contribution to $X(3823)\to J/\psi\pi^+\pi^-$ from the three Feynman diagrams in Fig. \ref{feynman} is
\begin{eqnarray}
\mathcal{M}_\mathrm{Total}&=&\mathcal{M}^{(a)}+e^{i\Phi}[\mathcal{M}^{(b)}+\mathcal{M}^{(c)}],\label{eqn17}
\end{eqnarray}
where we introduce the phase angle $\Phi$ as a measure of the interference between the amplitudes of the triangle and direct diagrams.
According to the three-body decay formula \cite{pdg2014}, the differential decay width for $X(3823)\to J/\psi\pi^+\pi^-$ is
\begin{eqnarray}
\mathrm{d}\Gamma=\frac{1}{(2\pi)^3}\frac{1}{32M_X^3}\overline{|{\mathcal{M}}_\mathrm{Total}|^2}\mathrm{d}m_{J/\psi\pi^+}^2\mathrm{d}m_{\pi^+\pi^-}^2, \label{eqn18}
\end{eqnarray}
where $m_{J/\psi\pi^+}^2=(p_4+k_1)^2$, and $m_{\pi^+\pi^-}^2=(k_1+k_2)^2$.\\

%\subsection{Numerical results and discussions}\label{sec3}
Before presenting our result, we need to determine the values of the coupling constants. The global coupling constant $g$ appearing in Eq.~(\ref{XHH}) can be obtained from Eqs.~\eqref{eqn6} and \eqref{psiDD} because we know the experimental value, $\Gamma[\psi(3770)\to D^0\bar{D}^0]=14.1$ MeV \cite{pdg2014}. The constant is thus $g=1.37\ \mathrm{GeV}^{-3/2}$. From Eq.~\eqref{XDDstar}, we obtain the coupling constant $g_{XDD^\ast}=12.7$.
Additionally, the coupling constant $g_{J/\psi DD}=8$ in Eq.~\eqref{eqn7} can be related to $g_{J/\psi DD^\ast}$, which is calculated using the vector meson dominance model and the QCD sum rule \cite{Deandrea:2003pv,Achasov:1994vh,Matheus:2002nq}.
Under heavy quark symmetry, we have the relation $g_{J/\psi DD}=g_{J/\psi D^\ast D^\ast}=m_D g_{J/\psi DD^\ast}$. Furthermore, $g_{\sigma DD}$ and $g_{\sigma D^\ast D^\ast}$ satisfy $g_{\sigma DD}=g_{\sigma D^\ast D^\ast}=m_{D^\ast}g_\pi/\sqrt{6}$ with $g_\pi=3.73$ \cite{Liu:2008xz,Bardeen:2003kt}. We use $m_\sigma=526$ MeV for the the mass of the $\sigma$ meson in our calculation. By fitting the decay width $\Gamma(\sigma\to \pi^+\pi^-)=200$ MeV \cite{Komada:2001jg}, we obtain the coupling constant $g_{\sigma\pi\pi}=1.8$ GeV. These values are summarized in Table \ref{tab:1}.

%Ref. \cite{Liu:2006df} has shown that the decay width in Eq. (\ref{eqn17}) is sensitive to $\alpha$ when $\alpha$ is chosen to be less than $3$, and

\begin{figure}[hptb]
	\scalebox{0.95}{\includegraphics[width=\columnwidth]{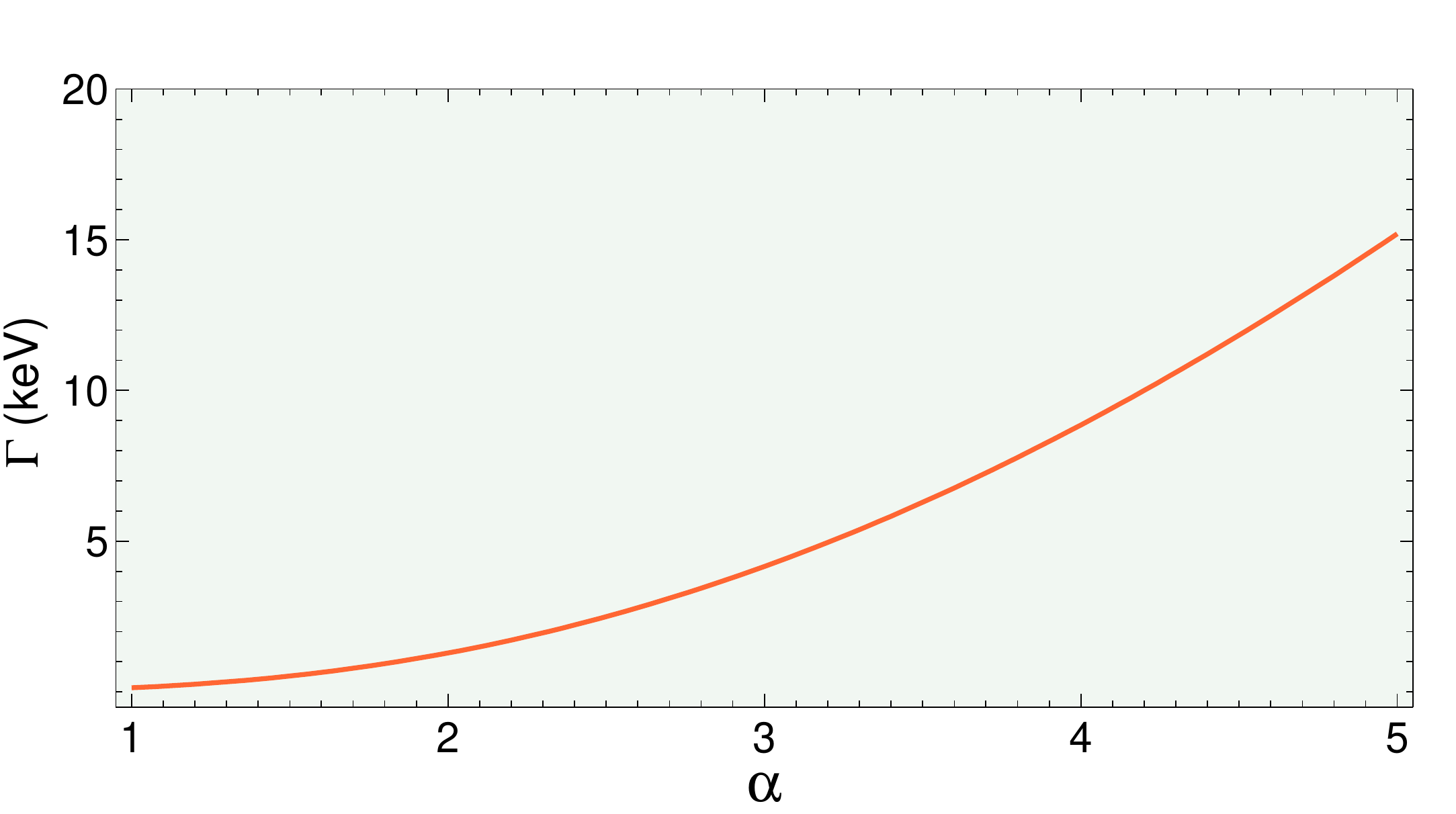}}
	\caption{(color online). Dependence of the decay width of $X(3823)\to DD^\ast\to J/\psi\pi^+\pi^-$ on $\alpha$ and phase angle $\Phi$ considering only Fig. \ref{feynman} (b) and (c).}\label{fig:5}
\end{figure}

With the above coupling constants as input, we first show the dependence on $\alpha$, in Eq.~\eqref{ff}, of the decay width of $X(3823)\to J/\psi\pi^+\pi^-$ if we consider only the contributions from Fig.~\ref{feynman} (b) and (c). The result, which is shown in Fig.~\ref{fig:5}, indicates that the contribution from the coupled-channel effect to $X(3823)\to J/\psi\pi^+\pi^-$ becomes larger when $\alpha$ is increased.
At present, we cannot fix the value of $\alpha$ using the experimental data. We choose a typical value of $\alpha=4.2$ as adopted in Ref.~\cite{Meng:2007cx}, where the decay width from the coupled-channel effect only is comparable with that from the direct decay.

\begin{table}[htbp]
	\caption{Values of the parameters used in our calculations.\label{tab:1}}
	\begin{center}
		\renewcommand{\arraystretch}{1.5}
		\tabcolsep=1.5pt
		\begin{tabular}{ccccc}
			\toprule[1pt]
			\toprule[1pt]
			$m_{X(3823)}$&                $m_{J/\psi}$&          $m_D$&                $m_{D^\ast}$&              $m_\pi$\\
			%\midrule[0.5pt]
			$3.823$ GeV&                $3.096$ GeV&           $1.865$ GeV&          $2.007$ GeV&               0.139 GeV\\
			\midrule[0.7pt]
			$g_{XJ/\psi\pi\pi}$&     $g_{XDD^\ast}$&         $g_{\psi DD}$&        $g_{J/\psi DD^\ast}$&      $g_{J/\psi DD}$\\
			%\midrule[0.5pt]
			$12.4$ GeV$^{-1}$&          $12.7$&                $-12.8$&              $4.3$ GeV$^{-1}$&                     $8$\\
			\midrule[0.7pt]
			$g_{\sigma DD}\ (g_{\sigma D^\ast D^\ast})$&  $g_{\sigma\pi\pi}$&  $m_\sigma$&    $\Gamma_\sigma$&     $\alpha$\\
			%\midrule[0.5pt]
			$3.1$ GeV                                 &$1.8$ GeV&               $0.526$ GeV&    $0.302$ GeV&        $4.2$\\
			\bottomrule[1pt]
			\bottomrule[1pt]
		\end{tabular}
	\end{center}
\end{table}

\section{Coupled-channel results}\label{sec4}

\begin{figure}[hptb]
	\scalebox{0.9}{\includegraphics[width=\columnwidth]{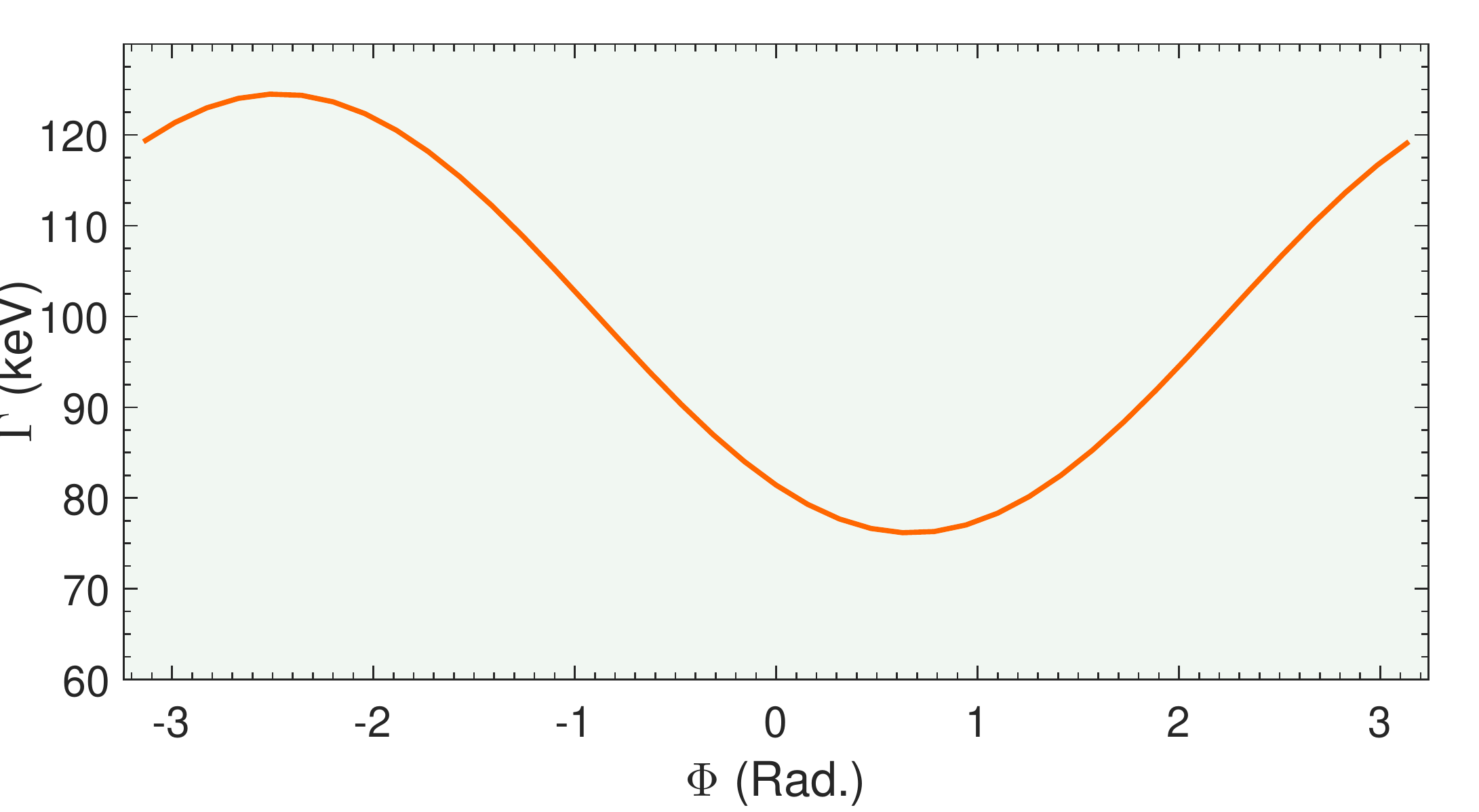}}
	\caption{(color online). Dependence of the total decay width of $X(3823)\to J/\psi\pi^+\pi^-$ on phase angle $\Phi$ considering Fig. \ref{feynman} (a), (b), and (c). Here, $\alpha$ was fixed at $4.2$ to obtain these results.}\label{fig:dependence}
\end{figure}

\begin{figure*}[htbp]
\setlength{\abovecaptionskip}{-0.1cm}
\centering
\subfigure
{
\begin{minipage}[b]{0.97\textwidth}
\includegraphics[width=1\textwidth]{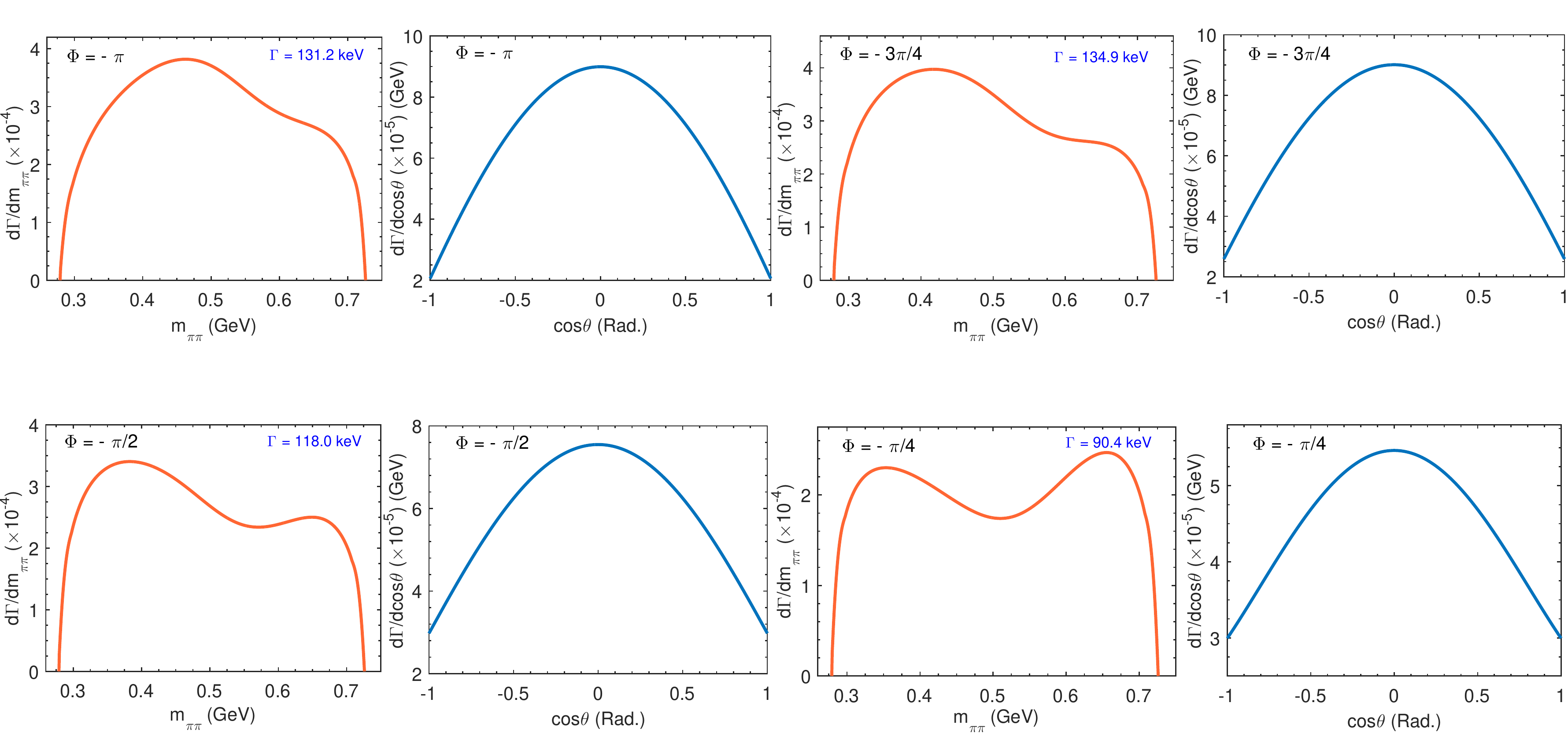} \\
\includegraphics[width=1\textwidth]{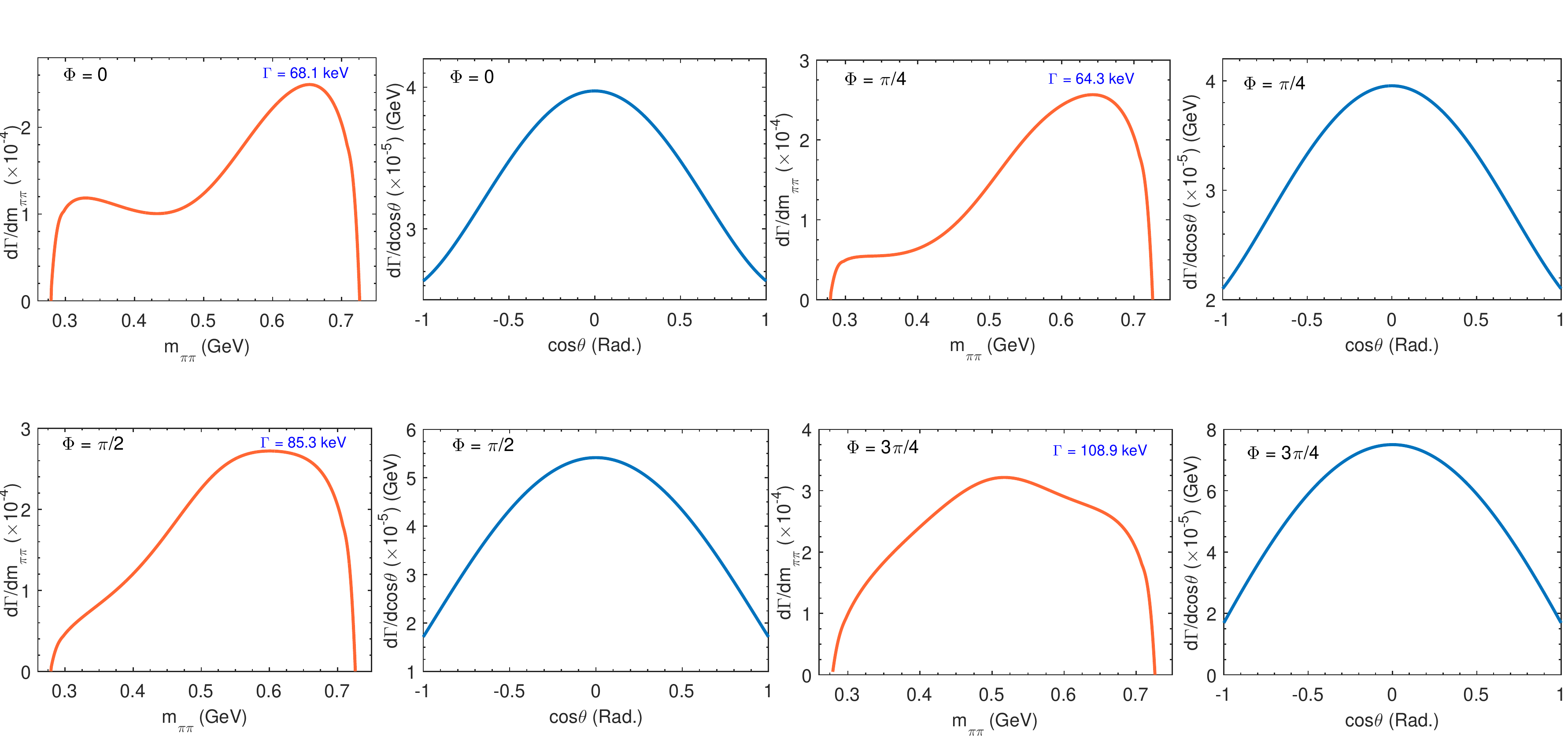}
\end{minipage}
}
\caption{The $\pi^+\pi^-$ invariant mass distribution $d\Gamma[X(3823)\to J/\psi\pi^+\pi^-]/dm_{\pi^+\pi^-}$ and the angular distribution $d\Gamma[X(3823)\to J/\psi\pi^+\pi^-]/d\cos\theta$ including the coupled-channel effect. Here, typical values of the phase angle $\Phi$ are taken, and the corresponding total decay width is listed.\label{fig:spectra}}
\end{figure*}

Now, we have fixed all the parameters in our work except for the phase angle $\Phi$ in Eq.~\eqref{eqn17}. Without any experimental constraints, this parameter is entirely free. Therefore, we compute the dependence of the total decay width of $X(3823)\to J/\psi\pi^+\pi^-$ on the phase angle $\Phi$, the $\pi^+\pi^-$ invariant mass distribution $\mathrm{d}\Gamma[X(3823)\to J/\psi\pi^+\pi^-]/\mathrm{d}m_{\pi^+\pi^-}$, and the polar angle distribution $\mathrm{d}\Gamma[X(3823)\to J/\psi\pi^+\pi^-]/\mathrm{d}\cos\theta$ for $\Phi$ values ranging over the entire trigonometric circle. These results are presented in Fig.~\ref{fig:dependence} and Fig.~\ref{fig:spectra}. The total decay width and the $\pi^+\pi^-$ invariant mass distribution are found to vary dramatically with $\Phi$, but the polar angle distributions almost keep the fixed line-shape.  For $\Phi$ within the third and fourth quadrants, the S-wave part is dominant, and most events scatter perpendicularly to the initial particle momentum, although the particular shape of the distribution changes slightly with the variation of $\Phi$. However, when $\Phi$ lies in the first and second quadrant, the contribution from S-wave is largely suppressed and D-wave becomes the dominant one, the angular distribution stays the same with when $\Phi$ lies in the third and fourth quadrant. Except for the cases $\Phi=3\pi/4$, it is difficult to distinguish which partial wave contribution is bigger in the $\pi^+\pi^-$ propagation, there seems to be a symmetric relation between the dominance of different partial wave and the value of $\Phi$. %When events are peaked at S-wave part, the $\Phi$ lies in the first and second quadrant, and when most events are peaked at D-wave part, the $\Phi$ lies in the third and fourth quadrant.
Fig.~\ref{fig:spectra} shows that we can't qualitatively judge the relative phase $\Phi$ from angle distribution, thus the precise measurement about the $\pi^+\pi^-$ mass spectrum is needed. %{\color{red} Fix here ---- EE.  Does not seem to agree with your Fig 5 where the scattering is clearly peaked at $\cos(\theta) = 0$ }%There is no symmetry in both distributions with the phase angle $\Phi$.

When we compare this result with the result from QCDME in Fig.~\ref{qcdspectrum} (a), we see there is always a difference between the $m_{\pi^+\pi^-}$ distributions with and without coupled-channel interference, for any angle $\Phi$, although for small $\Phi$ angles, the results become more similar. In addition, when we compare the results in Fig.~\ref{fig:spectra} with the sparse data in  Fig.~\ref{qcdspectrum} (b), we see that small angles for $\Phi$ are favored. Finally, the full widths, also given in Fig.~\ref{fig:spectra}, are sensitive to $\Phi$ as well. Given the analysis, we conclude that only data, so far sparse or nonexistent for $X(3823)\to J/\psi\pi^+\pi^-$, can yield more definite conclusions.\\

%=====================================================================================

\section{Summary and conclusions}\label{sec5}

Very recently, a true charmonium-like state, ~$X(3823)$, was established by the Belle and BESIII collaborations in the radiative decay channel $\chi_{c1}\gamma$. Moreover, a signal with low statistics was found two decades ago, by the E705 experiment in the same energy region, but in the hadronic decay channel $J/\psi\pi^+\pi^-$. This new state is most likely to be the missing charmonium $\psi(1^3D_2)$.

In this work, we studied the decay $X(3823)\to J/\psi\pi^+\pi^-$ within two different theoretical frameworks, the QCDME and the effective Lagrangian approach, including hadronic loops. The first method accounts for only the direct process, whereas the second method includes, not only the direct process, but also the coupled-channel effects due to the nearby closed OZI-allowed channel $D\bar{D}^*$. We computed the partial decay width distribution with the dipion invariant mass, $\Gamma[X(3823)\to J/\psi\pi^+\pi^-]/dm_{\pi^+\pi^-}$, within both approaches. If we neglect the coupled-channel effect and only consider the direct decay $X(3823)\to J/\psi\pi^+\pi^-$, within QCDME, we find disagreement with the scarce data from the E705 experiment, in particular for lower values of the dipion kinetic energy. This type of inaccuracy of the QCDME has been discussed in other works \cite{Kuang:2006me}, as the method does not include nonperturbative effects. If we include the coupled-channel by using the hadronic loop mechanism, we see there is always interference between the direct process and the indirect processes in Fig.~\ref{feynman}. The measure of the interference is an unknown phase $\Phi$. We vary $\Phi$ over its whole range and draw some conclusions: 1) The results of two methods do not match for any angle $\Phi$, showing that there is always a nonperturbative interference caused by the coupled-channel. 2) the specific line shape of the dipion invariant mass changes dramatically with $\Phi$, but the E705 data excludes values between $[-\pi,-\pi/4]$ and $[\pi/4,3\pi/4]$. 3) For small $\Phi$ angles, the decay distribution over the scattering angle $\theta$ is mostly in the perpendicular direction in relation with the momentum of the $X(3823)$. 4) Scenarios in Fig.~\ref{fig:spectra} corresponding to $\Phi$ between $[0,\pi/4]$ are favored. 5) Favored scenarios correspond to partial decay widths $\Gamma(X(3823)\to J/\psi\pi^+\pi^-)$ between 68.1 and 64.3 keV, values much within the experimental upper limit of 16 MeV.

Motivated by our nonperturbative coupled-channel results, we suggest new experimental studies in channel $X(3823)\to J/\psi\pi^+\pi^-$, which should be analyzed in terms of the $m_{\pi^+\pi^-}$ invariant mass distribution, and in the scattering angle distribution as well. We stress that such an analysis in this golden channel is relevant, not only to establish the expected hadronic decay of $\psi(1^3D_2)$, but will also provide theoretical insight into the contribution of coupled-channel effects in hadronic transitions.  Indeed, the present work already shows that the OZI-allowed channel, although it is closed, yet relatively nearby, influences an OZI-suppressed decay, however only experiment will allow us to quantify the strength of this influence and determine the $\Phi$ parameter. Nevertheless, we can already conclude that any realistic description of any hadron state should not neglect the nearby OZI-allowed hadronic decay channels. If this is true for the present state, which is not radially excited and is still below any opened channels, it will be even more true for the higher radially excited resonances or for light-quark systems.

\vfil
\section*{Acknowledgments}

This project is supported by the National Natural Science Foundation of China under Grant Nos.~11222547, 11175073, 11375240, and 11035006; the Ministry of Education of China (SRFDP under Grant No.~2012021111000); and the Chinese Academy of Sciences under the funding Y104160YQ0 and agreement No.~2015-BH-02.  This work is also supported by Fermilab, operated by the Fermi Research Alliance, LLC, U.S. Department of Energy, Contract DE-AC02-07CH11359 (EE).  %{\color{red} Note change --EE}


\begin{thebibliography}{199}

%\cite{Swanson:2006st} 1
\bibitem{Swanson:2006st}
  E.~S.~Swanson,
  %``The New heavy mesons: A Status report,''
  \href{http://www.sciencedirect.com/science/article/pii/S0370157306001475}{Phys.\ Rept.\  {\bf 429}, 243 (2006)}.


%\cite{Zhu:2007wz} 2
\bibitem{Zhu:2007wz}
  S.~L.~Zhu,
  %``New hadron states,''
  \href{http://arxiv.org/abs/hep-ph/0703225}{Int.\ J.\ Mod.\ Phys.\ E {\bf 17}, 283 (2008)}.

% 3
\bibitem{Brambilla:2010cs}
  N.~Brambilla, S.~Eidelman, B.~K.~Heltsley, R.~Vogt, G.~T.~Bodwin, E.~Eichten, A.~D.~Frawley and A.~B.~Meyer {\it et al.},
  %``Heavy quarkonium: progress, puzzles, and opportunities,''
  \href{http://link.springer.com/article/10.1140%2Fepjc%2Fs10052-010-1534-9}{Eur.\ Phys.\ J.\ C {\bf 71}, 1534 (2011)}.


  %\cite{Liu:2013waa} 4
\bibitem{Liu:2013waa}
  X.~Liu,
  %``An overview of $XYZ$ new particles,''
  \href{http://arxiv.org/abs/1312.7408}{Chin.\ Sci.\ Bull.\  {\bf 59}, 3815 (2014)}.

%\cite{Olsen:2014qna}
\bibitem{Olsen:2014qna}
  S.~L.~Olsen,
  %``A New Hadron Spectroscopy,''
  \href{http://link.springer.com/article/10.1007%2Fs11467-014-0449-6}{Front.\ Phys.\  {\bf 10}, 101401 (2015)}.
  %doi:10.1007/S11467-014-0449-6}.
 % [arXiv:1411.7738 [hep-ex]].
  %%CITATION = doi:10.1007/S11467-014-0449-6;%%
  %23 citations counted in INSPIRE as of 25 Dec 2015

%\cite{Yuan:2015kya}
\bibitem{Yuan:2015kya}
  C.~Z.~Yuan,
  %``Study of the XYZ states at the BESIII,''
  \href{http://link.springer.com/article/10.1007%2Fs11467-015-0484-y}{Front.\ Phys. {\bf 10}, 101401 (2015)}
  %doi:10.1007/s11467-015-0484-y}.
%  [arXiv:1509.06850 [hep-ex]].
  %%CITATION = doi:10.1007/s11467-015-0484-y;%%
  %1 citations counted in INSPIRE as of 25 Dec 2015

% 5
\bibitem{pdg2014}
  J.~Beringer {\it et al.}  [Particle Data Group Collaboration],
  %``Review of Particle Physics (RPP),''
  \href{http://pdg.lbl.gov/2014/tables/contents_tables_mesons.html}{Chins.\ Phys.\ C {\bf 38}, 090001 (2014)}.

  %\cite{Godfrey:1985xj} 6
\bibitem{Godfrey:1985xj}
  S.~Godfrey and N.~Isgur,
  %``Mesons in a relativized quark model with chromodynamics,''
  \href{http://journals.aps.org/prd/abstract/10.1103/PhysRevD.32.189}{Phys.\ Rev.\ D {\bf 32}, 189 (1985)}.
  %%CITATION = PHRVA,D32,189;%%
  %2036 citations counted in INSPIRE as of 22 May 2015

%\cite{Eichten:1978tg} 7
\bibitem{Eichten:1978tg}
  E.~Eichten, K.~Gottfried, T.~Kinoshita, K.~D.~Lane and T.~M.~Yan,
  %``Charmonium: The model,''
  \href{http://journals.aps.org/prd/abstract/10.1103/PhysRevD.17.3090}{Phys.\ Rev.\ D {\bf 17}, 3090 (1978)
  [Erratum-ibid.\ D {\bf 21}, 313 (1980)]}.
  %%CITATION = PHRVA,D17,3090;%%
  %1096 citations counted in INSPIRE as of 31 Mar 2015

%\cite{Antoniazzi:1993jz} 8
\bibitem{Antoniazzi:1993jz}
  L.~Antoniazzi {\it et al.}  [The E705 Collab.],
  %``Search for hidden charm resonance states decaying into $J/\psi$ or $\psi^\prime$ plus pions,''
  \href{http://journals.aps.org/prd/abstract/10.1103/PhysRevD.50.4258}{Phys.\ Rev.\ D {\bf 50}, 4258 (1994)}.
  %%CITATION = PHRVA,D50,4258;%%
  %45 citations counted in INSPIRE as of 31 Mar 2015

%\cite{Bhardwaj:2013rmw} 9
\bibitem{Bhardwaj:2013rmw}
  V.~Bhardwaj {\it et al.}  [The Belle Collab.],
  %``Evidence of a new narrow resonance decaying to $\chi_{c1}\gamma$ in $B \to \chi_{c1} \gamma K$,''
  \href{http://arxiv.org/abs/1304.3975}{Phys.\ Rev.\ Lett.\  {\bf 111}, no. 3, 032001 (2013)}.
  %%CITATION = ARXIV:1304.3975;%%
  %17 citations counted in INSPIRE as of 18 May 2015

  %\cite{Ablikim:2015baa} 10
\bibitem{Ablikim:2015baa}
  M.~Ablikim {\it et al.} [The BESIII Collab.]
  %``Observation of the $\psi(1^3D_2)$ state in $e^+e^-\to\pi^+\pi^-\gamma\chi_{c1}$ at BESIII,''
    \href{http://journals.aps.org/prl/abstract/10.1103/PhysRevLett.115.011803}{Phys.\ Rev.\ Lett. {\bf 115}, 011803 (2015)}.
%  \href{http://arxiv.org/abs/1503.08203}{arXiv:1503.08203}.
  %%CITATION = ARXIV:1503.08203;%%

%\cite{Eichten:2002qv} 11
\bibitem{Eichten:2002qv}
  E.~J.~Eichten, K.~Lane and C.~Quigg,
  %``$B$ meson gateways to missing charmonium levels,''
  \href{http://arxiv.org/abs/hep-ph/0206018}{Phys.\ Rev.\ Lett.\  {\bf 89}, 162002 (2002)}.
  %%CITATION = HEP-PH/0206018;%%
  %111 citations counted in INSPIRE as of 31 Mar 2015

%\cite{Cho:1994qp}
%\bibitem{Cho:1994qp}
%  P.~L.~Cho and M.~B.~Wise,
  %``Gluon fragmentation to D wave quarkonia,''
%  \href{http://journals.aps.org/prd/abstract/10.1103/PhysRevD.51.3352}{Phys.\ Rev.\ D {\bf 51}, 3352 (1995)}.
  %%CITATION = HEP-PH/9410214;%%
  %21 citations counted in INSPIRE as of 31 Mar 2015



%\cite{Ebert:2002pp} 12
\bibitem{Ebert:2002pp}
  D.~Ebert, R.~N.~Faustov and V.~O.~Galkin,
  %``Properties of heavy quarkonia and $B_c$ mesons in the relativistic quark model,''
  \href{http://journals.aps.org/prd/abstract/10.1103/PhysRevD.67.014027}{Phys.\ Rev.\ D {\bf 67}, 014027 (2003)}.
  %%CITATION = HEP-PH/0210381;%%
  %253 citations counted in INSPIRE as of 31 Mar 2015

%\cite{Ko:1997rn} 13
\bibitem{Ko:1997rn}
  P.~w.~Ko, J.~Lee and H.~S.~Song,
  %``Color octet mechanism in the inclusive $D$ wave charmonium productions in B decays,''
  \href{http://arxiv.org/abs/hep-ph/9701235}{Phys.\ Lett.\ B {\bf 395}, 107 (1997)}.
  %%CITATION = HEP-PH/9701235;%%
  %28 citations counted in INSPIRE as of 31 Mar 2015

%\cite{Qiao:1996ve} 14
\bibitem{Qiao:1996ve}
  C.~F.~Qiao, F.~Yuan and K.~T.~Chao,
  %``A Crucial test for color octet production mechanism in $Z^0$ decays,''
  \href{http://journals.aps.org/prd/abstract/10.1103/PhysRevD.55.4001}{Phys.\ Rev.\ D {\bf 55}, 4001 (1997)}.
  %%CITATION = HEP-PH/9609284;%%
  %16 citations counted in INSPIRE as of 31 Mar 2015

%\cite{Eichten:1994gt}
%\bibitem{Eichten:1994gt}
%  E.~J.~Eichten and C.~Quigg,
  %``Mesons with beauty and charm: Spectroscopy,''
%  \href{http://journals.aps.org/prd/abstract/10.1103/PhysRevD.49.5845}{Phys.\ Rev.\ D {\bf 49}, 5845 (1994)}.
  %[hep-ph/9402210].
  %%CITATION = HEP-PH/9402210;%%
  %357 citations counted in INSPIRE as of 31 Mar 2015

%\cite{Kuang:2006me} 15
\bibitem{Kuang:2006me}
  Y.~P.~Kuang,
  %``QCD multipole expansion and hadronic transitions in heavy quarkonium systems,''
  \href{http://arxiv.org/abs/hep-ph/0601044}{Front.\ Phys.\ China {\bf 1}, 19 (2006)}.
  %[hep-ph/0601044].
  %%CITATION = HEP-PH/0601044;%%
  %59 citations counted in INSPIRE as of 31 Mar 2015


%\cite{Liu:2006dq} 16
\bibitem{Liu:2006dq}
  X.~Liu, X.~Q.~Zeng and X.~Q.~Li,
  %``Study on contributions of hadronic loops to decays of $J/\psi\to vector + pseudoscalar$ mesons,''
  \href{http://journals.aps.org/prd/abstract/10.1103/PhysRevD.74.074003}{Phys.\ Rev.\ D {\bf 74}, 074003 (2006)}.
  %[hep-ph/0606191].
  %%CITATION = HEP-PH/0606191;%%
  %44 citations counted in INSPIRE as of 31 Mar 2015

%\cite{Simonov:2007bm} 17
\bibitem{Simonov:2007bm}
  Y.~A.~Simonov,
  %``Di-pion decays of heavy quarkonium in the field correlator method,''
  \href{http://arxiv.org/abs/0711.3626}{Phys.\ Atom.\ Nucl.\  {\bf 71}, 1048 (2008)}.
  %[arXiv:0711.3626 [hep-ph]].
  %%CITATION = ARXIV:0711.3626;%%
  %24 citations counted in INSPIRE as of 31 Mar 2015

%\cite{Liu:2009dr} 18
\bibitem{Liu:2009dr}
  X.~Liu, B.~Zhang and X.~Q.~Li,
  %``The Puzzle of excessive non-$D\bar{D}$ component of the inclusive $\psi(3770)$ decay and the long-distant contribution,''
  \href{http://arxiv.org/abs/0902.0480}{Phys.\ Lett.\ B {\bf 675}, 441 (2009)}.
  %[arXiv:0902.0480 [hep-ph]].
  %%CITATION = ARXIV:0902.0480;%%
  %36 citations counted in INSPIRE as of 31 Mar 2015

%\cite{Zhang:2009kr} 19
\bibitem{Zhang:2009kr}
  Y.~J.~Zhang, G.~Li and Q.~Zhao,
  %``Towards a dynamical understanding of the non-$D\bar{D}$ decay of $\psi(3770)$,''
  \href{http://journals.aps.org/prl/abstract/10.1103/PhysRevLett.102.172001}{Phys.\ Rev.\ Lett.\  {\bf 102}, 172001 (2009)}.
  %[arXiv:0902.1300 [hep-ph]].
  %%CITATION = ARXIV:0902.1300;%%
  %42 citations counted in INSPIRE as of 31 Mar 2015

%\cite{Chen:2011jp} 20
\bibitem{Chen:2011jp}
  D.~Y.~Chen, X.~Liu and X.~Q.~Li,
  %``Anomalous dipion invariant mass distribution of the $\Upsilon(4S)$ decays into $\Upsilon(1S) \pi^{+} \pi^{-}$ and $\Upsilon(2S) \pi^{+} \pi^{-}$,''
  \href{http://link.springer.com/article/10.1140%2Fepjc%2Fs10052-011-1808-x}{Eur.\ Phys.\ J.\ C {\bf 71}, 1808 (2011)}.
  %[arXiv:1109.1406 [hep-ph]].
  %%CITATION = ARXIV:1109.1406;%%

%\cite{Chen:2014ccr} 21
\bibitem{Chen:2014ccr}
  D.~Y.~Chen, X.~Liu and T.~Matsuki,
  %``Explaining the anomalous $\Upsilon(5S)\to \chi_{bJ}\omega$ decays through the hadronic loop effect,''
  \href{http://journals.aps.org/prd/abstract/10.1103/PhysRevD.90.034019}{Phys.\ Rev.\ D {\bf 90}, 034019 (2014)}.
  %[arXiv:1406.6763 [hep-ph]].
  %%CITATION = ARXIV:1406.6763;%%
  %2 citations counted in INSPIRE as of 31 Mar 2015

%\cite{Kuang:1988bz} 22
\bibitem{Kuang:1988bz}
  Y.~P.~Kuang, S.~F.~Tuan and T.~M.~Yan,
  %``Hadronic transitions and $P$ wave singlet states of heavy quarkonia,''
  \href{http://journals.aps.org/prd/abstract/10.1103/PhysRevD.37.1210}{Phys.\ Rev.\ D {\bf 37}, 1210 (1988)}.
  %%CITATION = PHRVA,D37,1210;%%
  %118 citations counted in INSPIRE as of 31 Mar 2015

% 23
\bibitem{Rupp:2015}
  G.~Rupp, E.~van Beveren, S.~Coito
  % "No serious meson spectroscopy without scattering"
  \href{http://www.actaphys.uj.edu.pl/sup8/abs/s8p0139}{Acta Phys.~Polon.~Suppl.~{\bf8}, 139 (2015)}.

%\cite{Yan:1980uh} 24
\bibitem{Yan:1980uh}
  T.~M.~Yan,
  %``Hadronic transitions between heavy quark states in quantum chromodynamics,''
  \href{http://journals.aps.org/prd/abstract/10.1103/PhysRevD.22.1652}{Phys.\ Rev.\ D {\bf 22}, 1652 (1980)}.
  %%CITATION = PHRVA,D22,1652;%%
  %262 citations counted in INSPIRE as of 18 May 2015

  %\cite{Kuang:1981se} 25
\bibitem{Kuang:1981se}
  Y.~P.~Kuang and T.~M.~Yan,
  %``Predictions for hadronic transitions in the $B\bar{B}$ system,''
  \href{http://journals.aps.org/prd/abstract/10.1103/PhysRevD.24.2874}{Phys.\ Rev.\ D {\bf 24}, 2874 (1981)}.
  %%CITATION = PHRVA,D24,2874;%%
  %236 citations counted in INSPIRE as of 18 May 2015

  %\cite{Buchmuller:1979gy}26
\bibitem{Buchmuller:1979gy}
  W.~Buchmuller and S.~H.~H.~Tye,
  %``Vibrational states in the $\Upsilon$ spectroscopy,''
  \href{http://journals.aps.org/prl/abstract/10.1103/PhysRevLett.44.850}{Phys.\ Rev.\ Lett.\  {\bf 44}, 850 (1980)}.
  %%CITATION = PRLTA,44,850;%%
  %48 citations counted in INSPIRE as of 27 juil. 2015

    %\cite{Voloshin:2015jua} 30
\bibitem{Voloshin:2015jua}
  M.~B.~Voloshin,
  %``The process $e^+e^- \to \pi \pi X(3823)$ in the soft pion limit,''
  \href{http://journals.aps.org/prd/abstract/10.1103/PhysRevD.91.114029}{Phys.\ Rev.\ D {\bf 91}, 114029 (2015)}.
%  \href{http://arxiv.org/abs/1504.02973v2}{arXiv:1504.02973}.
  %%CITATION = ARXIV:1504.02973;%%



   %\cite{Chen:2011zv} 26
\bibitem{Chen:2011zv}
  D.~Y.~Chen, X.~Liu and S.~L.~Zhu,
  %``Charged bottomonium-like states $Z_b(10610)$ and $Z_b(10650)$ and the $\Upsilon(5S)\to \Upsilon(2S)\pi^+\pi^-$ decay,''
  \href{http://journals.aps.org/prd/abstract/10.1103/PhysRevD.84.074016}{Phys.\ Rev.\ D {\bf 84}, 074016 (2011)}.
  %[arXiv:1105.5193 [hep-ph]].
  %%CITATION = ARXIV:1105.5193;%%
  %34 citations counted in INSPIRE as of 19 May 2015

  %\cite{Chen:2011qx} 27
\bibitem{Chen:2011qx}
  D.~Y.~Chen, J.~He, X.~Q.~Li and X.~Liu,
  %``Dipion invariant mass distribution of the anomalous $\Upsilon(1S) \pi^{+} \pi^{-}$ and $\Upsilon(2S) \pi^{+} \pi^{-}$ production near the peak of $\Upsilon(10860)$,''
  \href{http://journals.aps.org/prd/abstract/10.1103/PhysRevD.84.074006}{Phys.\ Rev.\ D {\bf 84}, 074006 (2011)}.
  %[arXiv:1105.1672 [hep-ph]].
  %%CITATION = ARXIV:1105.1672;%%
  %8 citations counted in INSPIRE as of 19 May 2015

  %\cite{Chen:2011xk} 28
\bibitem{Chen:2011xk}
  D.~Y.~Chen and X.~Liu,
  %``Predicted charged charmonium-like structures in the hidden-charm dipion decay of higher charmonia,''
  \href{http://journals.aps.org/prd/abstract/10.1103/PhysRevD.84.034032}{Phys.\ Rev.\ D {\bf 84}, 034032 (2011)}.
  %[arXiv:1106.5290 [hep-ph]].
  %%CITATION = ARXIV:1106.5290;%%
  %27 citations counted in INSPIRE as of 19 May 2015

 %\cite{Meng:2007cx} 29
\bibitem{Meng:2007cx}
  C.~Meng and K.~T.~Chao,
  %``Decays of the $X(3872) $ and chi(c1) (2P) charmonium,''
  \href{http://journals.aps.org/prd/abstract/10.1103/PhysRevD.75.114002}{Phys.\ Rev.\ D {\bf 75}, 114002 (2007)}.
  %[hep-ph/0703205].
  %%CITATION = HEP-PH/0703205;%%
  %57 citations counted in INSPIRE as of 19 May 2015

%\cite{Voloshin:1980zf}
\bibitem{Voloshin:1980zf}
  M.~B.~Voloshin and V.~I.~Zakharov,
  %``Measuring QCD anomalies in hadronic transitions between quarkonium states,''
  \href{http://journals.aps.org/prl/abstract/10.1103/PhysRevLett.45.688}{Phys.\ Rev.\ Lett.\  {\bf 45}, 688 (1980)}.
  %%CITATION = PRLTA,45,688;%%
  %239 citations counted in INSPIRE as of 14 Aug 2015

  %\cite{Voloshin:2006ce} 31
\bibitem{Voloshin:2006ce}
  M.~B.~Voloshin,
  %``Two-pion transitions in quarkonium revisited,''
  \href{http://journals.aps.org/prd/abstract/10.1103/PhysRevD.74.054022}{Phys.\ Rev.\ D {\bf 74}, 054022 (2006)}.
  %[hep-ph/0606258].
  %%CITATION = HEP-PH/0606258;%%
  %26 citations counted in INSPIRE as of 18 May 2015

    %\cite{Casalbuoni:1996pg} 32
\bibitem{Casalbuoni:1996pg}
  R.~Casalbuoni, A.~Deandrea, N.~Di Bartolomeo, R.~Gatto, F.~Feruglio and G.~Nardulli,
  %``Phenomenology of heavy meson chiral Lagrangians,''
  \href{http://arxiv.org/abs/hep-ph/9605342v2}{Phys.\ Rept.\  {\bf 281}, 145 (1997)}.
  %[hep-ph/9605342].
  %%CITATION = HEP-PH/9605342;%%
  %387 citations counted in INSPIRE as of 19 May 2015

   %\cite{Casalbuoni:1992fd} 33
\bibitem{Casalbuoni:1992fd}
  R.~Casalbuoni, A.~Deandrea, N.~Di Bartolomeo, R.~Gatto, F.~Feruglio and G.~Nardulli,
  %``Hadronic transitions among quarkonium states in a soft exchange approximation. Chiral breaking and spin symmetry breaking processes,''
  \href{http://arxiv.org/abs/hep-ph/9304280v1}{Phys.\ Lett.\ B {\bf 309}, 163 (1993)}.
  %[hep-ph/9304280].
  %%CITATION = HEP-PH/9304280;%%
  %24 citations counted in INSPIRE as of 19 May 2015

  %\cite{Colangelo:2003sa} 34
\bibitem{Colangelo:2003sa}
  P.~Colangelo, F.~De Fazio and T.~N.~Pham,
  %``Nonfactorizable contributions in $B$ decays to charmonium: The case of $B\to K^- h_c$,''
  \href{http://journals.aps.org/prd/abstract/10.1103/PhysRevD.69.054023}{Phys.\ Rev.\ D {\bf 69}, 054023 (2004)}.
  %[hep-ph/0310084].
  %%CITATION = HEP-PH/0310084;%%
  %94 citations counted in INSPIRE as of 20 May 2015

    %\cite{Oh:2000qr} 35
%\bibitem{Oh:2000qr}
%  Y.~s.~Oh, T.~Song and S.~H.~Lee,
  %``J / psi absorption by pi and rho mesons in meson exchange model with anomalous parity interactions,''
%  \href{http://journals.aps.org/prc/abstract/10.1103/PhysRevC.63.034901}{Phys.\ Rev.\ C {\bf 63}, 034901 (2001)}.
  %[nucl-th/0010064].
  %%CITATION = NUCL-TH/0010064;%%
  %123 citations counted in INSPIRE as of 20 May 2015

    %\cite{Liu:2006df} 35
\bibitem{Liu:2006df}
  X.~Liu, B.~Zhang and S.~L.~Zhu,
  %``The hidden charm decay of $X(3872)$, $Y(3940)$ and final state interaction effects,''
  \href{http://www.sciencedirect.com/science/article/pii/S0370269306015747}{Phys.\ Lett.\ B {\bf 645}, 185 (2007)}.
  %[hep-ph/0610278].
  %%CITATION = HEP-PH/0610278;%%
  %42 citations counted in INSPIRE as of 21 May 2015

  %\cite{Chen:2013coa} 36
\bibitem{Chen:2013coa}
  D.~Y.~Chen, X.~Liu and T.~Matsuki,
  %``Reproducing the $Z_c(3900)$ structure through the initial-single-pion-emission mechanism,''
  \href{http://journals.aps.org/prd/abstract/10.1103/PhysRevD.88.036008}{Phys.\ Rev.\ D {\bf 88}, no. 3, 036008 (2013)}.
  %[arXiv:1304.5845 [hep-ph]].
  %%CITATION = ARXIV:1304.5845;%%
  %31 citations counted in INSPIRE as of 21 May 2015


  %\cite{Deandrea:2003pv} 37
\bibitem{Deandrea:2003pv}
  A.~Deandrea, G.~Nardulli and A.~D.~Polosa,
  %``$J/\psi$ couplings to charmed resonances and to $\pi$,''
  \href{http://journals.aps.org/prd/abstract/10.1103/PhysRevD.68.034002}{Phys.\ Rev.\ D {\bf 68}, 034002 (2003)}.
  %[hep-ph/0302273].
  %%CITATION = HEP-PH/0302273;%%
  %48 citations counted in INSPIRE as of 21 May 2015

  %\cite{Achasov:1994vh} 38
\bibitem{Achasov:1994vh}
  N.~N.~Achasov and A.~A.~Kozhevnikov,
  %``Dynamical violation of the OZI rule and G parity in the decays of heavy quarkonia,''
  \href{http://journals.aps.org/prd/abstract/10.1103/PhysRevD.49.275}{Phys.\ Rev.\ D {\bf 49}, 275 (1994)}.
  %%CITATION = PHRVA,D49,275;%%
  %21 citations counted in INSPIRE as of 21 May 2015

  %\cite{Matheus:2002nq} 39
\bibitem{Matheus:2002nq}
  R.~D.~Matheus, F.~S.~Navarra, M.~Nielsen and R.~Rodrigues da Silva
  %``The $J/\psi D D$ vertex in QCD sum rules,''
  \href{http://www.sciencedirect.com/science/article/pii/S0370269302022591}{Phys.\ Lett.\ B {\bf 541}, 265 (2002)}.
  %[hep-ph/0206198].
  %%CITATION = HEP-PH/0206198;%%
  %67 citations counted in INSPIRE as of 21 May 2015

   %\cite{Liu:2008xz} 40
\bibitem{Liu:2008xz}
  X.~Liu, Y.~R.~Liu, W.~Z.~Deng and S.~L.~Zhu,
  %``$Z^+(4430)$ as a $D_1^\prime D^* (D_1 D^*)$ molecular state,''
  \href{http://journals.aps.org/prd/abstract/10.1103/PhysRevD.77.094015}{Phys.\ Rev.\ D {\bf 77}, 094015 (2008)}.
  %[arXiv:0803.1295 [hep-ph]].
  %%CITATION = ARXIV:0803.1295;%%
  %55 citations counted in INSPIRE as of 21 May 2015
  %\cite{Liu:2008xz}

%\cite{Bardeen:2003kt} 41
\bibitem{Bardeen:2003kt}
  W.~A.~Bardeen, E.~J.~Eichten and C.~T.~Hill,
  %``Chiral multiplets of heavy - light mesons,''
  \href{http://journals.aps.org/prd/abstract/10.1103/PhysRevD.68.054024}{Phys.\ Rev.\ D {\bf 68}, 054024 (2003)}.
  %[hep-ph/0305049].
  %%CITATION = HEP-PH/0305049;%%
  %374 citations counted in INSPIRE as of 21 May 2015

    %\cite{Komada:2001jg} 42
\bibitem{Komada:2001jg}
  T.~Komada, S.~Ishida and M.~Ishida,
  %``The sigma meson production in excited Upsilon decay processes: Phenomenological analyses,''
  \href{http://www.sciencedirect.com/science/article/pii/S0370269301004051}{Phys.\ Lett.\ B {\bf 508}, 31 (2001)}.
  %%CITATION = PHLTA,B508,31;%%
  %33 citations counted in INSPIRE as of 21 May 2015

    %\cite{Colangelo:2002mj}
%\bibitem{Colangelo:2002mj}
%  P.~Colangelo, F.~De Fazio and T.~N.~Pham,
  %``B- ---> K- (chi(c0)) decay from charmed meson rescattering,''
%  \href{http://www.sciencedirect.com/science/article/pii/S0370269302023067}{Phys.\ Lett.\ B {\bf 542}, 71 (2002)}.
  %[hep-ph/0207061].
  %%CITATION = HEP-PH/0207061;%%
  %79 citations counted in INSPIRE as of 21 May 2015


  %\cite{Meng:2007tk}
%\bibitem{Meng:2007tk}
%  C.~Meng and K.~T.~Chao,
  %``Scalar resonance contributions to the dipion transition rates of Upsilon(4S,5S) in the re-scattering model,''
%  \href{http://journals.aps.org/prd/abstract/10.1103/PhysRevD.77.074003}{Phys.\ Rev.\ D {\bf 77}, 074003 (2008)}.
  %[arXiv:0712.3595 [hep-ph]].
  %%CITATION = ARXIV:0712.3595;%%
  %40 citations counted in INSPIRE as of 22 May 2015


  %\cite{Kuang:1989ub}
%\bibitem{Kuang:1989ub}
%  Y.~P.~Kuang and T.~M.~Yan,
  %``Hadronic Transitions of $D$ Wave Quarkonium and $\psi(3770) \to J/\psi \pi \pi$,''
%  \href{http://journals.aps.org/prd/abstract/10.1103/PhysRevD.41.155}{Phys.\ Rev.\ D {\bf 41}, 155 (1990)}.
  %%CITATION = PHRVA,D41,155;%%
  %75 citations counted in INSPIRE as of 18 May 2015



  %\cite{Novikov:1980fa}
%\bibitem{Novikov:1980fa}
%  V.~A.~Novikov and M.~A.~Shifman,
  %``Comment on the psi-prime ---> J/psi pi pi Decay,''
%  \href{http://link.springer.com/article/10.1007%2FBF01429829}{Z.\ Phys.\ C {\bf 8}, 43 (1981)}.
%CITATION = ZEPYA,C8,43;%%
%167 citations counted in INSPIRE as of 22 May 2015

%\cite{Voloshin:1980zf}
%\bibitem{Voloshin:1980zf}
%  M.~B.~Voloshin and V.~I.~Zakharov,
  %``Measuring QCD Anomalies in Hadronic Transitions Between Onium States,''
% \href{http://journals.aps.org/prl/abstract/10.1103/PhysRevLett.45.688}{ Phys.\ Rev.\ Lett.\  {\bf 45}, 688 (1980)}.
  %%CITATION = PRLTA,45,688;%%
  %235 citations counted in INSPIRE as of 22 May 2015







  %\cite{Neubert:1993mb}
%\bibitem{Neubert:1993mb}
%  M.~Neubert,
  %``Heavy quark symmetry,''
%  \href{http://www.sciencedirect.com/science/article/pii/0370157394900914}{Phys.\ Rept.\  {\bf 245}, 259 (1994)}.
  %[hep-ph/9306320].
  %%CITATION = HEP-PH/9306320;%%
  %1248 citations counted in INSPIRE as of 19 May 2015













\end{thebibliography}
\end{document}